\documentclass[floatfix,twocolumn]{aastex7}
\usepackage{amsmath}
\usepackage{bm}
\usepackage[normalem]{ulem}
\usepackage{wrapfig}
\usepackage{hyperref}
\newcommand*\dif{\mathop{}\!\mathrm{d}}

\received{April 22, 2025}
\revised{May 9, 2025}
\accepted{May 17, 2025}

\submitjournal{ApJ}

\shorttitle{Line and Delay Profiles in NGC 5548 Motivate a Multicomponent BLR}
\shortauthors{Long et al.}

\graphicspath{{./}{figures/}}

\begin{document}

\title{\vspace{-8mm}Reverberation Mapping Data of NGC 5548 Imply a Multicomponent Broad-Line Region}

\author{Kirk Long}
\affiliation{JILA, University of Colorado Boulder}
\email{kirk.long@colorado.edu}
\correspondingauthor{Kirk Long}
\email{kirk.long@colorado.edu}
\author{Jason Dexter}
\affiliation{JILA, University of Colorado Boulder}
\email{jason.dexter@colorado.edu}

\begin{abstract}
Line broadening and variability are observational hallmarks of active galactic nuclei which allow us to measure supermassive black hole masses as well as constrain the geometry and kinematics of the emitting gas, with the most precise measurements requiring a degree of modeling. Two popular models of the broad-line region describe the emitting gas as either a distribution of puffed up clouds or a thin disk with strong velocity gradients. As we show in this work, key features in the reverberation mapping dataset obtained by the AGN STORM team in NGC 5548 cannot be accounted for by either simple model. In several emission lines the observed broad-line profile has a single peak yet the delay profile has a distinct double peak, strongly motivating a BLR with emission from multiple components. We demonstrate a few possibilities that may alleviate the tension and better represent the true nature of the broad-line emitting gas in NGC 5548 and beyond.
\end{abstract}

\section{Introduction}
Quasars and active galactic nuclei (AGN) exhibit significant line broadening, an observable feature that, in conjunction with a radial distance estimate from the central supermassive black hole (SMBH), allows us to infer the mass of the SMBH. This broadening implies the gas is moving at high velocity due to the central black hole's gravity \citep{Peterson2006}. Assuming the gas in the broad-line region is virialized, we can write a simple relationship between the observables and black hole mass:
\begin{equation}\label{RMeqn}
    M_{BH} = f \frac{R(\Delta V)^2}{G},
\end{equation}

The characteristic velocity of the line $\Delta V$ can be measured from the width of the broad-emission line in a single epoch spectrum, but measuring $R$\textemdash the characteristic distance from the black hole to the emitting gas\textemdash is more difficult. $R$ can be estimated with either reverberation mapping (RM) \citep{BLANDFORD_MCKEE_82,Peterson93_RM} or inteferometric \citep{GRAVITY+,GRAVITY17,GRAVITY19} techniques. RM relies on substituting temporal resolution for spatial resolution, while the interferometric measurement can at present be done only with the GRAVITY instrument on the VLTI for a select group of bright, relatively nearby quasars. While both methods require expensive, dedicated observing campaigns, and can therefore not feasibly be completed for the entire population of broad-line AGN, they are critical as they calibrate many further SMBH mass measurements. Observations from both techniques show there is a radius-luminosity (R-L) relation $R\propto L^\alpha$ for the BLR, enabling single-epoch mass estimations for sources without GRAVITY or RM observations \citep{Bentz2013_Hbeta_R-L,Halpha_R-L_Cho_2023,GRAVITY_R-L_2024,2024arXivDallaBonta}. 

Finally, $f$ is the so-called ``virial factor", which is of order unity and contains all the unknown kinematics of the BLR. While $f$ can be estimated empirically, there is thus a certain degree of systematic model-dependent uncertainty in estimating SMBH masses. The most precise mass measurements from either GRAVITY or RM observations rely on forward modeling different schematic pictures for the geometry and kinematics of the BLR, thus it is critical we understand what models best fit the data and the systematic errors associated with our choice of model \citep{Long_2023}.

In this work we will focus primarily on the reverberation mapping technique, which measures lags between continuum and emission line changes by taking many spectra over a long time baseline, observing how the line profile changes over time. The observed delays between continuum and emission line changes can be used to measure $R\approx c\Delta t$\textemdash assuming such changes propagate at the speed of light $c$ and that the continuum is point-like (from the perspective of the BLR) and originates near the black hole \citep{Peterson2006,BLANDFORD_MCKEE_82,RM_Review_21,Peterson2014_REVIEW,Peterson93_RM}. Formally, reverberation mapping assumes variations in the emission line are related to variations in the continuum via a transfer function $\Psi$:

\begin{equation}\label{transfer}
    L(v,t) = \int \Psi(v,\tau)C(t-\tau)\mathrm{d}\tau
\end{equation}
Where $L$ is the emission line and $C$ the continuum. $\Psi$ encodes the geometry, kinematics, and radiative transfer effects of the BLR. Reverberation mapping has been successfully used to estimate the masses of many AGN in the local universe, and both RM and GRAVITY measurments have demonstrated that the BLR is usually within a few light-days (or several thousand Schwarszchild radii) of the central SMBH, making studying the BLR a uniquely direct way to study these objects outside our galaxy. 

A common approach when fitting models of the BLR to reverberation mapping data is to match the line profile and produce a synthetic lightcurve to compare to observations, either across the entire emission line (1D) or as a function of velocity (2D). While early RM fits were largely 1D only \citep{memecho2}, the remarkable data quality produced by campaigns such as AGN-STORM have allowed for 2D fitting of the BLR and the recovery of velocity-delay maps either through inversion methods \citep{Grier_2013_ECHO_INFALL,STORM_ECHO} or model-fitting \citep{STORM_CLOUDS}. Here we will focus on focus on the viability of various models in fitting the 2D reverberation mapping data products, in particular focusing on the velocity-resolved line and delay profiles published for NGC 5548 in \cite{STORM_MODEL_SPECTRA}.

Specifically, we demonstrate how and why two current popular models for the BLR fail to satisfactorily explain the full data products, in particular the lag/delay profiles, and present several possible schematics of more complicated BLR geometries that could alleviate this problem. We focus our discussion solely on the AGN-STORM reverberation mapping dataset of NGCC 5548 \citep{STORM_PROP,STORM_MODEL_SPECTRA,STORM_CCF}, specifically in the Lyman $\alpha$ line, but note that it is not unique to this line or this source, as similar behavior has been observed both in other lines of NGC 5548 and in completely different sources \citep{STORM2_II,RM3c27319}. 

\section{Summary of BLR models and implementation}
\label{sec:models}

Before discussing the problems with fitting two popular models of the BLR to the data taken by the AGN-STORM team, we first provide a brief summary of the ``disk-wind" and ``cloud" models that will be referred to throughout this work. The disk-wind model we use is an expanded version of the line optically thick Sobolev \citep{Sobolev} disk-wind model originally presented by \cite{CM96} which we have also used in our previous work fitting GRAVITY data to the BLR in 3C 273 \citep{Long_2023}. In the line optically thick limit the line emission is modified such that photons from particular locations in the disk have a much larger escape probability $p_{\rm{esc}}$, and this escape probability is formally related to the line of sight velocity gradients \textbf{($\lvert\frac{\rm{d}v_l}{\rm{d}l}\rvert$)} in the disk \citep{RH83}. 

In our implementation of the disk-wind model we describe this \citep[following][]{CM96} as: 
\begin{equation} \label{eq6}
\begin{split}
p_{\rm{esc}} & \propto \left\lvert\frac{\mathrm{d}v_l}{\mathrm{d}l}\right\rvert \approx \left\lvert\hat{n}\cdot\boldsymbol{\Lambda}\cdot\hat{n}\right\rvert \\& \approx \bigg\lvert\sqrt{\frac{1}{2r^3}} \bigg (3\sin^2i\cos\phi\bigg[\sqrt{2}{f_1}\cos\phi + {f_2}\frac{\sin\phi}{2}\bigg]\\& - 3 {f_3} \sin i \cos i \cos\phi + \sqrt{2}{f_4} \cos^2i\bigg)\bigg\rvert
\end{split}
\end{equation} 
\textbf{$\left\lvert\hat{n}\cdot\boldsymbol{\Lambda}\cdot\hat{n}\right\rvert $ encodes how the fluid effects from the rate of strain tensor $\Lambda$ are seen by the observer, and is calculated following \cite{Batchelor68}, with $f_{1,2,3,4}$ representing possible shears in the disk. The disk is geometrically thin and its orientation is such that $\phi = 0$ is along the line of sight and nearest the observer.} A full derivation of equation \ref{eq6} is available in \cite{Long_2023} (equations 5-8 and the Appendix), as well as a full description of the disk-wind model which in summary is described by the following parameters:
\begin{enumerate}
    \item The inclination angle of the system $i$, where $i = 90^\circ$ corresponds to an edge-on viewing angle and $i = 0^\circ$ a ``face-on" viewing angle. 
    \item The mass of the central supermassive black hole, $M_{\textrm{BH}}$. 
    \item The mean radius of the BLR $\hat{r}$ as weighted by the emissivity, i.e. 
    \begin{equation}
        \bar{r} = \frac{\int_{A} rj \dif A}{\int_{A} j \dif A} = \frac{\int\limits_{r_{\textrm{min}}}^{r_{\textrm{max}}} r^2 j(r) \mathrm{d}r}{\int\limits_{r_{\textrm{min}}}^{r_{\textrm{max}}} r j(r) \mathrm{d}r}
    \end{equation}\
    The emissivity $j(r)$ is described as a power-law with an intrinsic line source function $S(r)\propto r^{-\alpha}$. Higher values of $\bar{r}$ lead to the line/delay profiles being ``squeezed" in wavelength space.
    \item A size scaling factor $r_{\textrm{fac}}$, which in conjunction with $\bar{r}$ gives the minimum and maximum radii of the BLR via $r_{\textrm{min}} = \bar{r}\frac{(3-2\alpha)}{(1-2\alpha)}\left(\frac{r_{\textrm{fac}}^{1/2-\alpha} - 1}{r_{\textrm{fac}}^{3/2-\alpha}-1}\right)$ and $r_{\textrm{max}} = r_{\textrm{fac}} r_{\textrm{min}}$. Increasing $r_{\textrm{fac}}$ slightly ``stretches" the line/delay profiles in wavelength space. 
    \item The proportional strengths of the shear terms $f_{1,2,3,4}$, where $f_1$ controls the radial shear, ${f_2}$ the Keplerian shear, ${f_3}$ the radial lifting shear, and $f_4$ the height lifting shear.
    \item A parameter $n$ that can vary the normalization of the line profile with respect to the data slightly as the data points may not be exactly at the peak of the line, $n \ge 1$, where $n = 1$ corresponds to scaling the model exactly to the maximum flux measurement in the data.
    \item A parameter $\Delta \lambda_c$ that varies the line center, thus slightly shifting the models left and right in wavelength space.
\end{enumerate}

We also test and combine this disk-wind model with a thicker disk/cloud model similar to that described in \cite{Pancoast2011,PANCOAST_CLOUDS_14} given the popularity of such cloud models in the literature. In implementing the cloud model used in this work we follow the parametrization of \cite{PANCOAST_CLOUDS_14} and refer the interested reader to their work for a full description of this model and the motivations behind it. To summarize briefly, the Pancoast cloud model draws random point particles on roughly Keplerian orbits confined by an opening angle and radial distribution function but otherwise randomly distributed around the black hole. In addition to the inclination, mass, mean radius, $\Delta \lambda_c$, and $n$ parameters described above our simplified implementation of the Pancoast-style cloud model of the BLR as used in this work additionally uses the following parameters:
    \begin{enumerate}
        \item A term that describes the shape of the radial cloud distribution, $\beta$.
        \item A term that sets a minimum radius for the distribution of clouds, $F$, where $F$ is the fraction of $\mu$ that clouds must at a minimum be located.
        \item The opening angle of the distribution of clouds, $\theta_0$.
        \item A term that describes how closely concentrated the clouds are to the extent of the opening angle, $\gamma$.
        \item A term that describes how obscured the bottom half of the clouds are, $\xi$, where $\xi = 1.0$ corresponds to every cloud from beneath the midplane of the system having been removed and $\xi = 0.0$ corresponds to an equal distribution of clouds across the midplane.
        \item A term that describes anisotropy in azimuthal emission, $\kappa$, allowing for emission to be more concentrated from certain $\phi$ ranges within the cloud distribution (i.e. lighting up the front/back stronger than the sides or vice versa). 
    \end{enumerate}
There are further complications possible to this model, including turbulence, elliptical orbits, and inflow which we do not discuss in this work as we found they are not important for the main conclusions presented here.

There is a further hidden physical choice in both classes of models\textemdash when one considers the delays measured across the emission line how strongly a parcel of gas ``responds" is not required to be the same as how strongly it emits. Photoionization theory suggests the BLR gas responsivity $\eta$ may have a power law dependence, i.e. $\eta \propto r^{\beta}$ with $0\lesssim \beta \lesssim 1.5$ \citep{Korista_Goad_response_2004,Goad_Korista_response_2014}. For simplicity we keep a constant responsivity similar to previous modeling as we did not find that changing $\eta$ affected the results discussed here. 

We also explore three more complicated BLR geometries (case 1, 2, and 3 further detailed in section \ref{sec:cases}) in this work which add additional parameters to these models. In the first case we add two parameters to each model, a $\phi_{\rm{min}}$ and $\phi_{\rm{max}}$ that limit the angular region of each model that can emit, assuming the rest is arbitrarily blocked off by an ``obscurer". This addition results in a total of up to 12 free parameters in the disk-wind case or 13 free parameters in the cloud case. 

In our \hyperref[case2]{case 2} model, we add a simple thin-disk model to a \hyperref[case1]{case 1} cloud model, and we also allow for the velocity field of the disk to have an inflowing component in addition to a Keplerian component. This introduces a new parameter that varies the strength of the disk and cloud components against each other, as a well as a parameter that controls how much the velocity field is inflowing vs. purely rotating, but it discards the $f$ terms as parameters as the emission is assumed to be azimuthally isotropic. The black hole mass is fixed to be the same for both components but every other parameter is potentially independent, up to 21 free parameters.

Finally, in our \hyperref[case3]{case 3} model we consider a combination of superimposed cloud models. With two components, this consists of adding a \hyperref[case1]{case 1} cloud model to another \hyperref[case1]{case 1} cloud model, which again introduces a new parameter to fix the ratio of their intensities. Again the black hole mass is fixed between each model but every other parameter is potentially independent, up to 25 free parameters.

These models are implemented using the publicly available Julia package \texttt{BroadLineRegions.jl} and their use is fully documented in Long (2025, in prep, with current implementation available \href{https://github.com/kirklong/BroadLineRegions.jl}{here}).

\section{Simple BLR models fail to explain RM data in NGC 5548}\label{sec:problem}
The data in NGC 5548\textemdash in particular the delay profile\textemdash appear to imply a disk-like component \citep{STORM_ECHO,STORM_MODEL_SPECTRA}. The broad, single-peaked line profiles are hard to explain with simple disks, however, thus they are often explained by modeling the BLR as a distribution of clouds \citep{Pancoast2011,PANCOAST_CLOUDS_14} or with more complicated disk-wind type models \citep{RH83,CM96,CM97,Flohic2012,ChajetHall2013,Long_2023}. As we show here, for either the clouds or disk-wind models of the BLR one \textbf{cannot match both the delay and line profile observed}. 

\begin{figure*}[!ht]
    \begin{minipage}{\textwidth}
    \centering
    \includegraphics[width=1.\linewidth,keepaspectratio]{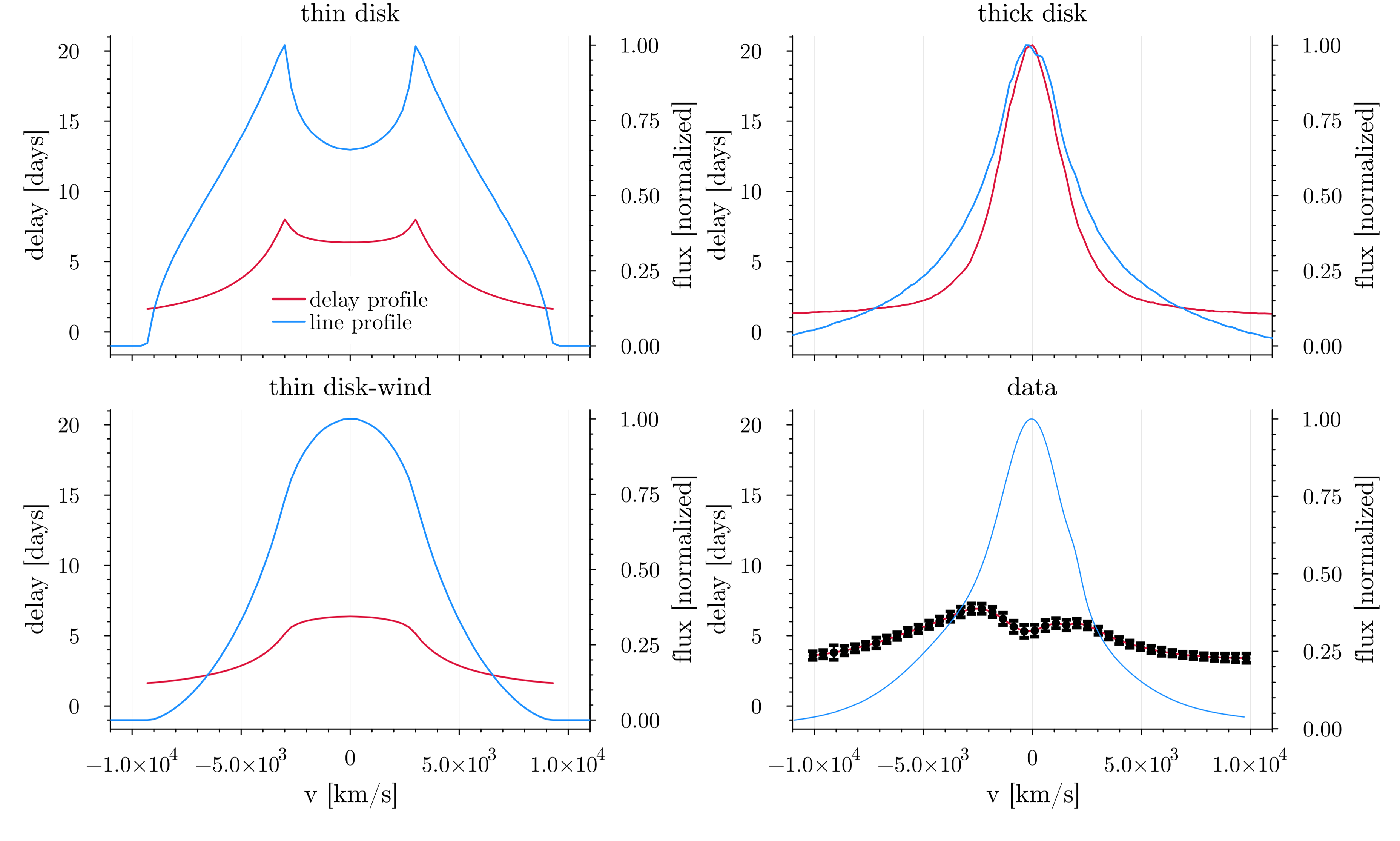}
    \caption{Here we show three sets of line and delay profiles for representative models discussed in this work, where each should be compared with data profiles shown in the bottom right panel. The ``thin disk" parameters were chosen to match the locations of the peaks in the delay profile seen in the Lyman $\alpha$ data, but notice that both the line profile and amplitude\textemdash the difference between the peak and wing delays\textemdash are wrong. The line profile data do not have visible error bars due to the smoothness of the measurements and robust HST calibration. The ``thin disk-wind" case has similar physical parameters to the ``thin disk" case, but uses the Sobolev approximation and radial shear such that the line profile and delay profile become single-peaked. The cloud (thick disk) profile example is chosen from parameters designed to match those presented in \cite{STORM_CLOUDS}, obtained from fitting the lightcurve and line profile data. Note that for all of the representative models the width of the delay profile is narrower than the line profile. While both the cloud and disk-wind models can have asymmetries in both the line and delay profiles, the models shown here are chosen for their simplicity/match to the data and thus do not show the asymmetry seen particularly in the delay profile data.}
    \label{fig:cloudSinglePeakDelay}
    \end{minipage}
\end{figure*}

To understand why this problem arises it is best to first consider why there are two peaks in the ``standard" azimuthally isotropic emitting disk under Keplerian rotation. \citet{Welsh&Horne_EchoMaps} showed that the peaks in the delay profile of a thin Keplerian disk map onto the outermost annulus of the disk, with the space between the peaks set by the difference in velocity between the red and blue sides of this outermost ring. If one imagines a ray in velocity-delay space going through the system at one of the peaks, it is tangent with the ellipse corresponding to the outer edge of the disk and thus picks up many more contributions from farther regions than inner regions of the disk, weighting the observed delay to be longer. If one imagines a similar ray going through the system near zero velocity, it is perpendicular to all ellipses in velocity-delay space and thus picks up more equal contributions from all areas of the system, resulting in the observed delay dropping in the line core. The wings are dominated by emission that is increasingly closer and closer to the center of the system, and thus their delays also drop with respect to these two peaks. 

In the thin disk-wind case where radial shear dominates the light is preferentially emitted from the front and back sides of the disk as a result of radiative transfer effects\textemdash this results in the observed emission that comes from where the two peaks should be to instead originate from gas closer to the system center but nearer $\phi \approx 0,\pi$ in the azimuthal plane, resulting in shorter delays than in the isotropically emitting case \citep{CM96}. Thus the longest possible observed delays in a radial shear dominated disk-wind model similar to \cite{CM96} are near where the velocity $v\approx0$, as this is the only area where the entire system is sampled. This results in single peaks in both the delay and line profiles.

But why does a thicker, puffed up cloud-like BLR have a single peak in the delay profile? When the clouds are puffed up and disordered their velocities are more randomly distributed along the line of sight when compared to the thin-disk case. This results in the modeled emission line profile obtaining a single peak that matches what is observed. Unfortunately, this also results in a single peak in the delay profile: because the clouds are ``disordered" when they are puffed up to larger heights, the most likely cloud to find at zero velocity is one far away and conversely the only cloud one can find at the highest velocities is nearest the center, assuming the clouds are drawn from a smooth radial distribution with random distributions in azimuthal angle $\phi$. Since they are then randomly distributed, there are no projection/viewing angle effects, and the delay map is only sensitive to this spatial distribution of the clouds, which necessarily peaks at zero velocity as the most distant clouds will most often reside there. While the cloud model does have parameters that can change the intensity weighting of the observed distribution\textemdash see $\xi$ and $\kappa$ in Section \ref{sec:models}\textemdash these weighting schemes preserve front/back symmetry and thus are not complicated enough to explain the data observations.

This behavior is shown in Figure \ref{fig:cloudSinglePeakDelay}. Note again that this is also partially due to the fact that the clouds are drawn from a smooth radial distribution with front-back symmetry and then randomly shuffled about in azimuth and tilt\textemdash one could imagine various contrived ways of placing clouds that could potentially alleviate this problem, but this is difficult to physically justify, and current modeling schemes do not cherry-pick the locations of the clouds to the degree required. While the disk-wind model has ways to break this symmetry through the incorporation of shears other than/in addition to radial shear, doing so correspondingly breaks the single-peak nature of the line profile, thus neither model can satisfactorily explain the observations.

\begin{figure*}[!ht]
    \begin{minipage}{\textwidth}
        \centering
        \includegraphics[width=1.\linewidth,keepaspectratio]{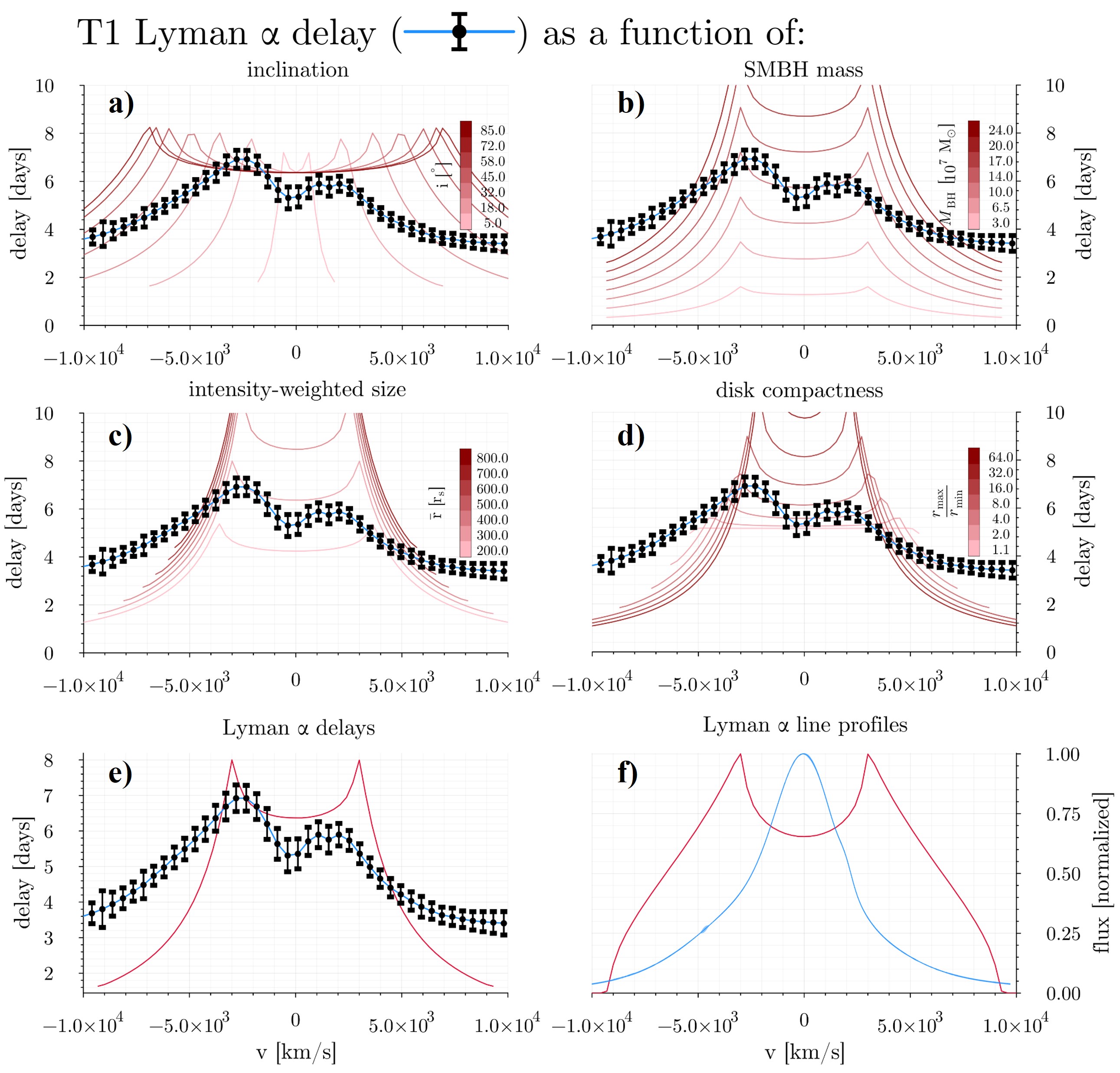}
        \caption{\textbf{a-d}: The data (here shown is just the ``T1" period identified by the STORM team, but note that a similar two peak structure and amplitude persists throughout the observations for the Lyman alpha line) appear to show a disk-like structure, but are hard to interpret through the framework of a simple disk due to the amplitude of the observed profile and the location of the two peaks. Altering the responsivity or considering a slightly more complicated disk model can soften the sharpness of the peaks \citep{HorneMarsh86_KeplerianShear}, but the difference in delay between the peaks and the wings is difficult to reconcile with any disk-type model. Shown here are four fundamental parameters for a simple thin-disk type model, demonstrating that although the structure appears ``disky" it cannot be explained by a thin-disk component alone. \newline
        \textbf{e-f}: Shown here are the base parameters used in creating panels a-d, chosen to match the velocity locations of the peaks and overall amplitude of the delay profile. Notice that they struggle to capture the wings of the data in the delay profile, and the line profile has both the wrong morphology and\textemdash more importantly\textemdash the wrong width. The line profile data is shown in blue as before but with a ribbon to illustrate the errors, which are barely visible at this scale, as velocity resolution in the line profile is much higher and the errors much smaller than the corresponding quantities in the delay profile. Also note that, as discussed in the main text, the simple disk models shown here do not incorporate any inflow and thus cannot capture the asymmetry shown in the delay profile. While adding radial inflow can change the heights between the two peaks, this does not resolve the broader problems outlined in this section.\vspace{10mm}}
        \label{fig:dataDelayProb}
    \end{minipage}
\end{figure*}

While the delay profile morphology appears to imply a disk-like structure \citep{STORM_MODEL_SPECTRA}, as shown in Figure \ref{fig:dataDelayProb}, even taken alone the delay profile is difficult to explain with a single disk type model. Although the delay profile data shows the characteristic double-peaked structure indicative of Keplerian disks, the amplitude\textemdash the difference between the delays at the peaks and in the wings\textemdash is inconsistent with what one might expect if the reverberating material were just a disk. One immediate discrepancy from a simple rotating disk is the fact that the delay profile is asymmetric\textemdash the negative velocity peak corresponds to longer delays than the positive velocity peak. This asymmetry could be the result of inflow or more complicated BLR effects \citep[][and references therein]{RM3c27319}, and will be discussed in more detail in Section \ref{sec:cases}, but in the following discussion we ignore this asymmetry and assume a simple, circularly rotating disk.

For Lyman alpha, the data show a maximum absolute velocity of $\sim \pm$ 10,000 km/s, which implies $r_{\rm{min}}\sim 500 r_s (\sin^2 i)$ from $v=\sqrt{\frac{GM}{r}}$. Assuming the system were a simple, circular disk allows us to estimate the maximum radius of the disk via the velocity locations of the peaks. The peaks are located $\sim 2000-2500$ km/s from line center, and this corresponds to a maximum velocity-implied size of $r_{\rm{max}}\sim 10,000 r_s (\sin^2 i)$. 

For a Keplerian disk the expected time delay is $t = \frac{r}{c}\left(1-\cos\phi\sin i\right)$ \citep{WATERS16}. The possible range of observations is thus $\frac{r}{c}\left(1-\sin i\right)<t<\frac{r}{c}\left(1+\sin i\right)$ with an average of $\frac{r}{c}$ for a complete ring. Thus the $\Delta t$ between complete rings at $r_{\rm{max}}$ and $r_{\rm{min}}$ can be estimated as:
\begin{equation}
\Delta t \sim \frac{r_{\rm{max}}-r_{\rm{min}}}{c} = \frac{\left (r_{\rm{max}}/r_{\rm{min}}-1\right )r_{\rm{min}}}{c}
\label{dataProblemTwiddleMath}
\end{equation}

The innermost gas should map directly to the minimum lag of $\sim 3.5$ days, while the delay at the peaks ($\sim 6-7$ days) should be heavily skewed towards sampling the largest radii of the system. The ratio between the inner and outer annuli of the disk as measured from the amplitude of the delays is then at least $\sim 2\sim\frac{r_{\rm{max}}}{r_{\rm{min}}}$, thus:

$$ \Delta t_{\rm{\tau}} \gtrsim \frac{r_{\rm{min}}}{c}$$
The value expected from inferring the location from the velocities alone is much higher, however, as the ratio of $r_{\rm{max}}/r_{\rm{min}}$ inferred from the velocity locations of the peak is roughly 15-25, implying: 

$$\Delta t_{\rm{v}} \sim 20\frac{r_{\rm{min}}}{c}$$ 
Even though we have employed simple arguments in arriving at these estimates, the discrepancy is clearly large enough $\left(\frac{\Delta t_{\rm{v}}}{\Delta t_{\rm{\tau}}} \lesssim 20\right)$ that a more complicated model is needed to explain the data, even without considering the line profile. While the width of the velocity bins and inclination effects can lead to some underestimation of $\Delta t_{\rm{\tau}}$, in simple disk models observed with similar resolution as the data we find this underestimation factor is at most $\sim 5$ and thus not enough to alleviate this tension $\left(\frac{\Delta t_v}{\Delta t_{\rm{\tau}}}\gtrsim4\right)$.

Note that one cannot simply resolve this problem by saying the farthest away gas is unconstrained by observations in an attempt to bring the amplitude of the peaks down. If this were the case we would expect to see a plateau-like feature in the delay profile capped at the maximum delay capable of being constrained by the observations, but the sharpness of the two resolved peaks rules this out, at least in the case of Lyman $\alpha$. Similarly, while the responsivity $\eta$ can somewhat affect the width and amplitude of the delay profile it does not change the overall morphology \citep[single vs. double peak; ][]{Li_2025_Response}. 

But there is not just the delay profile to contend with\textemdash the observed line profile is still broad and single-peaked in contrast to the double-peaked delay profile. As shown in Figure \ref{fig:dataDelayProb}, the closest class of thin-disk model parameters in the delay space produce a line profile that is much too broad in addition to necessarily having the wrong morphology (two peaks instead of one). Combined, the broad, single peak in the line profile in addition to the interesting problems in the delay profile noted above strongly motivate a different understanding of the BLR, at least in NGC 5548. 

In summary, the unambiguous detection of a double-peaked profile in the delays across the Ly$\alpha$ line in NGC 5548 by the AGN STORM campaign in conjunction with the broad, single-peaked line profile cannot be explained with popular single component BLR models. While the line profile can be accurately recreated with either a cloud or disk-wind type model, the delay profile cannot be reconstructed from either alone. This implies that either the responding gas that creates the delay profile is significantly decoupled from the emitting gas that creates the line profile, or\textemdash if one believes both structures in the data are signatures from the same gas\textemdash more complicated modeling is required. We will focus on the second possibility, of which we present a few possible solutions in the next section. 

\section{Sample, viable kinematic BLR models of NGC 5548}\label{sec:cases}
While the discussion below employs largely qualitative arguments, we would like to emphasize we have attempted more rigorous fitting of models to infer parameters and have determined that such an approach is beyond the scope of this paper, given the problems presented in the previous and next sections. Many of the qualitative arguments in this paper were born out of our attempts to robustly fit the data with many different model prescriptions for the BLR, and for the interested reader we further detail these fitting attempts, the problems, and the lessons learned in the \hyperref[appendix]{Appendix}. 

\textbf{Case 1:} \textit{Obscured single component models}\label{case1}\newline
    The problem\textemdash with either a disk-wind style BLR or puffed up cloudy BLR\textemdash is that current models produce a single peak in the delay profile when there is a single peak in the line profile due to how the emitting gas must be distributed. The delays near $v=0$ can be shortened, however, if one simply were to obscure some or all of the emission from the far side of the system and the velocities are relatively ordered (disk-like). 
    \begin{figure}[!ht]
        \begin{minipage}{0.45\textwidth}
            \centering
            \includegraphics[width=1.\linewidth,keepaspectratio]{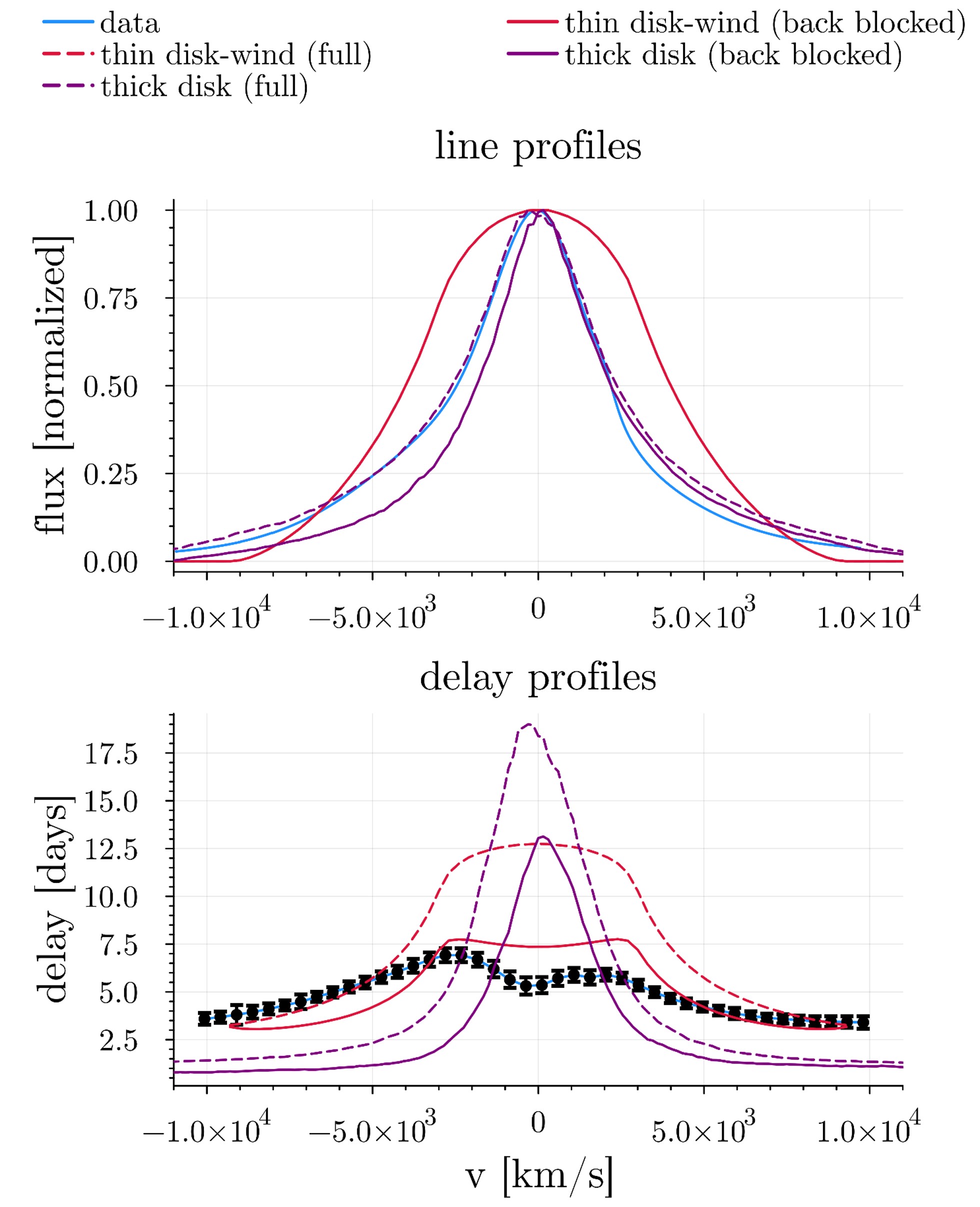}
            \caption{Sample disk-wind and cloud models, where the solid line in each case indicates the same model but with the far side obscured from the observer. Like in Figure \ref{fig:cloudSinglePeakDelay} the parameters for the cloud (thick disk) model are chosen to roughly match those previously fit to NGC 5548 in \cite{STORM_CLOUDS} and the disk-wind parameters were chosen to roughly match the scaling of the delay profile. Note that the line profile remains the same for the disk-wind case as it is perfectly symmetric front/back in intensity space, thus the normalized line profile does not change when blocking off the back. Also note that both models are chosen for simplicity and thus do not include effects that may be responsible for the asymmetry of the delay profile as discussed in the main text. The line and delay data are shown in blue, with black points on top of the delay measurements indicating the uncertainty of each measurement\textemdash the line profile data are so finely sampled and of such high quality that their errors are not visible on this scale.}
            \label{fig:case1}
        \end{minipage}
    \end{figure} 
    Since the system is roughly left-right symmetric the line profile shape is preserved and thus only the delay profile is affected. This behavior is easiest to produce in disk-type models where the velocities are ordered. An example profile from both a representative cloud and disk-wind type model is shown in Figure \ref{fig:case1}.
    
    In the case of the disk-wind model this can qualitatively explain the features of the data\textemdash the sample delay curve shown has two peaks in roughly the right locations and is roughly the right amplitude\textemdash but quantitatively the width of the line profile is much too large, thus there is still a problem between the implied scales of the system in the line profile and delay profile as detailed above. In the case of the clouds the line profile is better matched, but while blocking off the far side of the cloud distribution acts to shorten the delays it does not change the shape of the delay profile due to the disordered nature of the clouds. Thus while some single component models can match the location of the peaks, the central dip, and the line profile shape relatively well, we have not been able to find a solution that correctly matches the \textit{amplitude} of both the delay profile and the width of the line profile for the reasons further detailed in previous sections. This strongly motivates a two component model, which we discuss two possibilities for below: 

\textbf{Case 2:} \textit{Partially inflowing disk with clouds} \label{case2}\newline
    The obvious and most physically intuitive choice to explain the data is simply to invoke that the BLR has a disky, inflowing component and an extended, puffed up cloud component. As discussed above, disks can naturally match the general morphology of the delay data but the amplitude of the delay curve in NGC 5548 seems at odds with the width and shape of the line profile if it were to be caused by a simple Keplerian disk emitting roughly isotropically in azimuth. This can be ``fixed'' if there is a second, roughly level component in the delay space, which naturally shrinks the amplitude of the delay profile. 
    \begin{figure}[!ht]
        \begin{minipage}{0.45\textwidth}
        \centering
        \includegraphics[width=1.\linewidth,keepaspectratio]{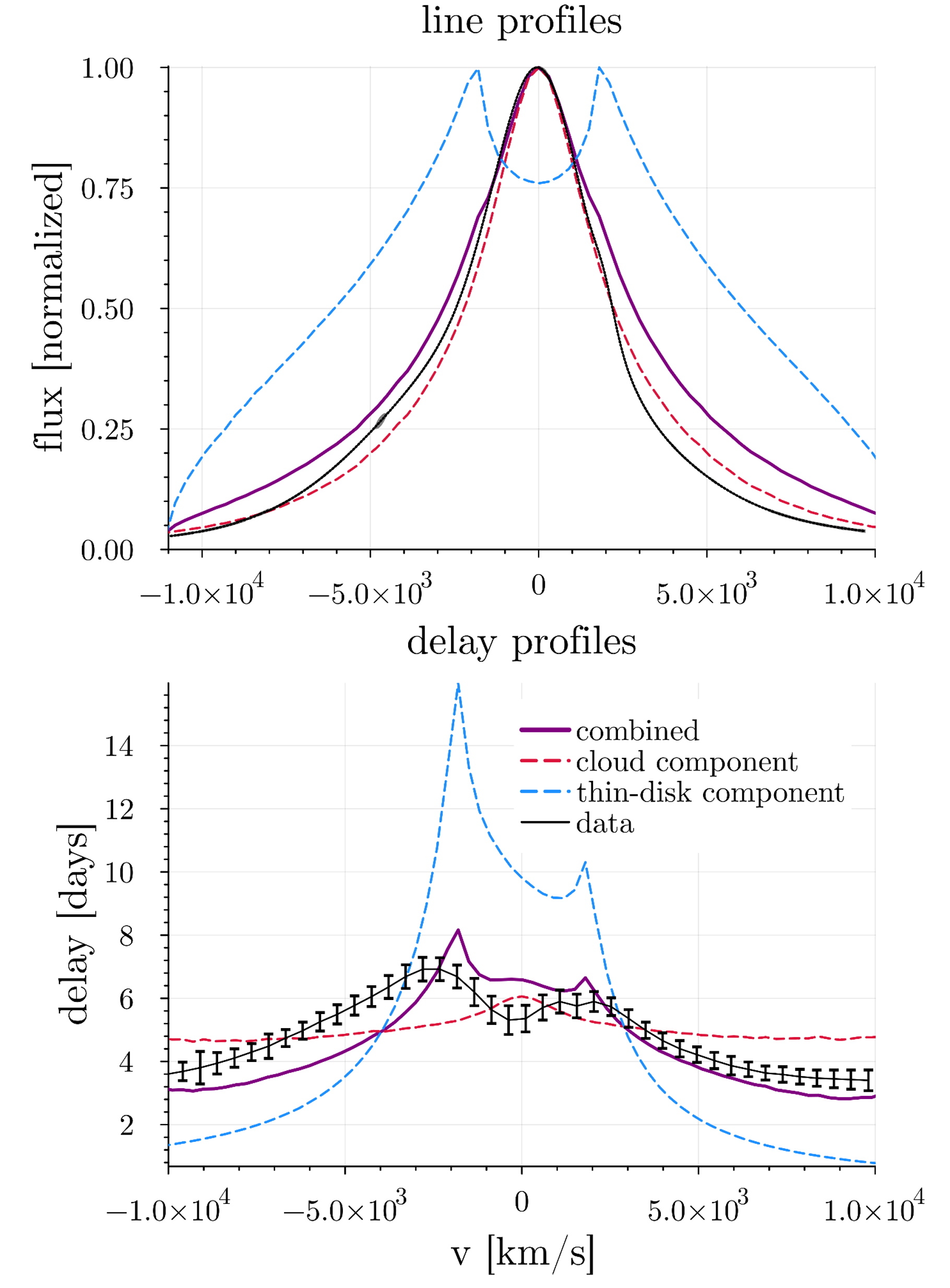}
        \caption{Sample geometry and profiles of a two component BLR, where the inner component is a thin disk emitting isotropically azimuthally and with a power-law drop-off in intensity radially, with a velocity field that is $2/3$ Keplerian and $1/3$ radial inflow. The second component is a distribution of clouds with an opening angle larger than the inclination angle and concentrated around a $\sim 5$ light-day radius such that they act to ``level" out the amplitude of the disk delay component and fill in the hole left by the disk component in the line profile. }
        \label{fig:case2}
        \end{minipage}
    \end{figure}
    This can be accomplished most easily by invoking a cloud distribution on top of a disk model, with the clouds having a relatively constrained radial extent to keep their delay profile mostly flat but an opening angle large enough such that the velocities are randomly distributed along the line of sight and thus the line profile has a single peak. 
    
    This means that, by number, most of the clouds are located closer to zero velocity, thus they dominate this region of the line and delay profile, with the disk component dominating the wings. Thus although the line profile of the disk component alone should have two peaks, the dominance of the cloud line profile in the region near zero velocity allows this component to ``fill in" the gap left by the disk, while the wings of the disk set the appropriate strengths nearest to the black hole. In the delay profile this same behavior acts to moderate the amplitude of the peaks from the disk profile\textemdash while alone the peaks would be much larger in delay space, the average of the peak height with the clouds in this region brings the amplitude in line with expectations while keeping the overall morphology correct. 
    
    Finally, noting that the delay profile appears to have a negative slope between the left and right velocity peaks, we add a radially inflowing component to the disk velocity to match this morphology. Note that this inflow component is implemented into the velocity field as a fixed fraction of the Keplerian rotational velocity, and we do not consider velocity gradients in this simple toy model\textemdash that is we do not use Sobolev approximation used in the full disk-wind model and instead keep the approximation of a thin, azimuthally isotropic emitting disk for simplicity here.
    
    Sample line and delay profiles for this prescription are shown in Figure \ref{fig:case2}. Unfortunately this is a very complicated toy model, with 21 free parameters as described in this work and possibly many more if turbulence/Sobolev effects were included, thus we do not attempt to formally fit the data and instead present this hand-picked picture as an illustrative possibility. 

\textbf{Case 3:} \textit{multicomponent cloud BLR} \label{case3} \newline
    \begin{figure}[!ht]
        \begin{minipage}{0.45\textwidth}
        \centering
        \includegraphics[width=1.\linewidth,keepaspectratio]{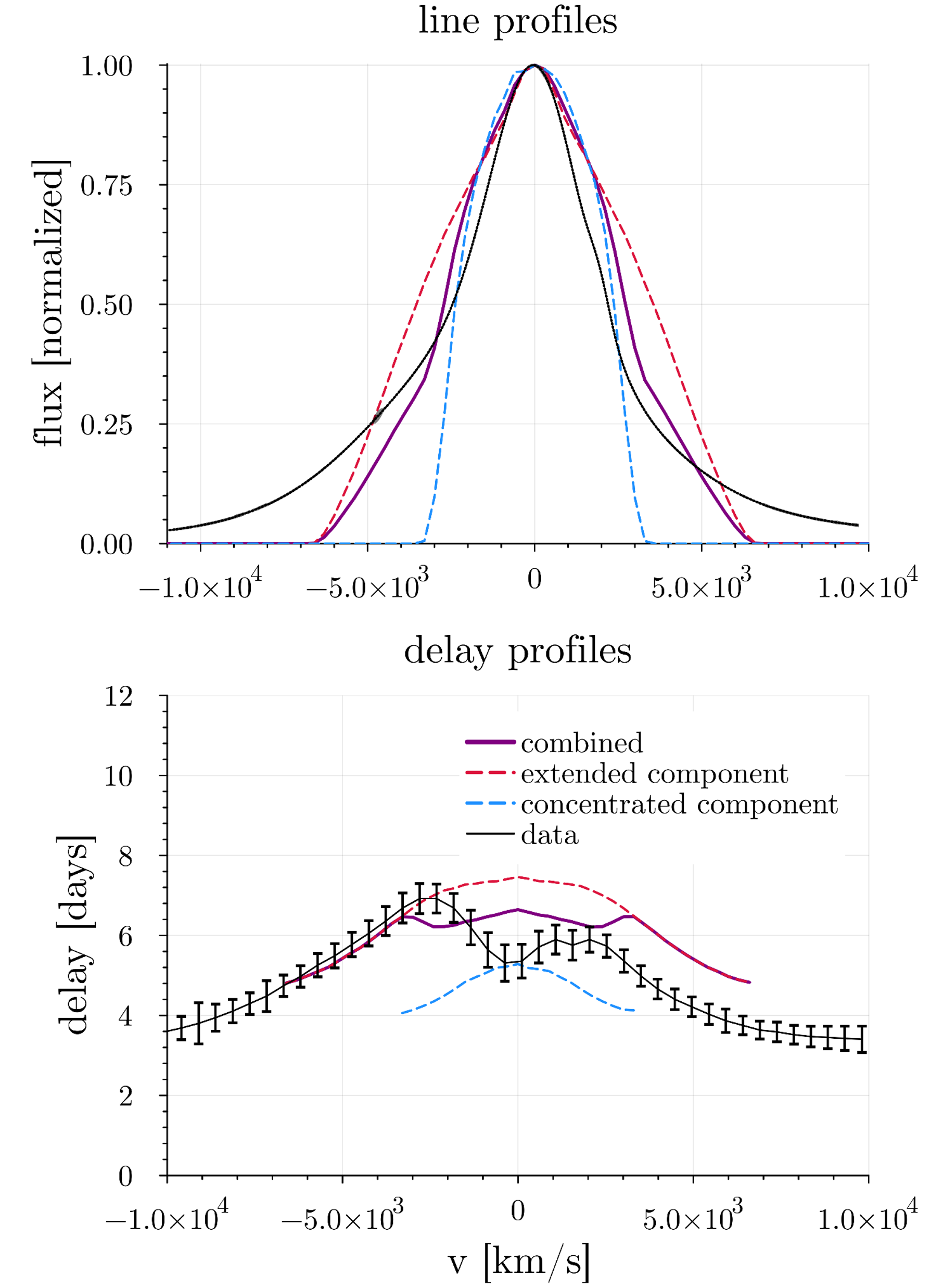}
        \caption{An example of the multiple cloud component BLR idea. Note that while this does not appear to match as well as the \hyperref[case2]{case 2} visualization, both were chosen by hand and their appearance here is only to illustrate the general idea and not to present a best fit of either model.}
        \label{fig:case3}
        \end{minipage}
    \end{figure}
    One interesting and more exotic possible explanation is that the double-peak in the delay profile is the result of two (or more) different distributions of emitting gas located at different distances from the black hole. In particular, if there were a narrower (in velocity space) width component with a single peaked line and delay profile, but with a small amplitude in delay space, and this component were underlying a broader single peaked line and delay profile component with a larger amplitude in delay space, such a configuration would easily produce two peaks in the delay profile while preserving a single peak in the line profile. In order for this picture to work one must also ensure that the inner distribution has a comparable or greater total responsivity than the outer distribution, such that near line center it dominates the delay profile and drags the center down to make the double-peaked structure. Sample line and delay profiles for this prescription are shown in Figure \ref{fig:case3}.

    \begin{figure}[!ht]
        \begin{minipage}{0.45\textwidth}
        \centering
        \includegraphics[width=1.\linewidth,keepaspectratio]{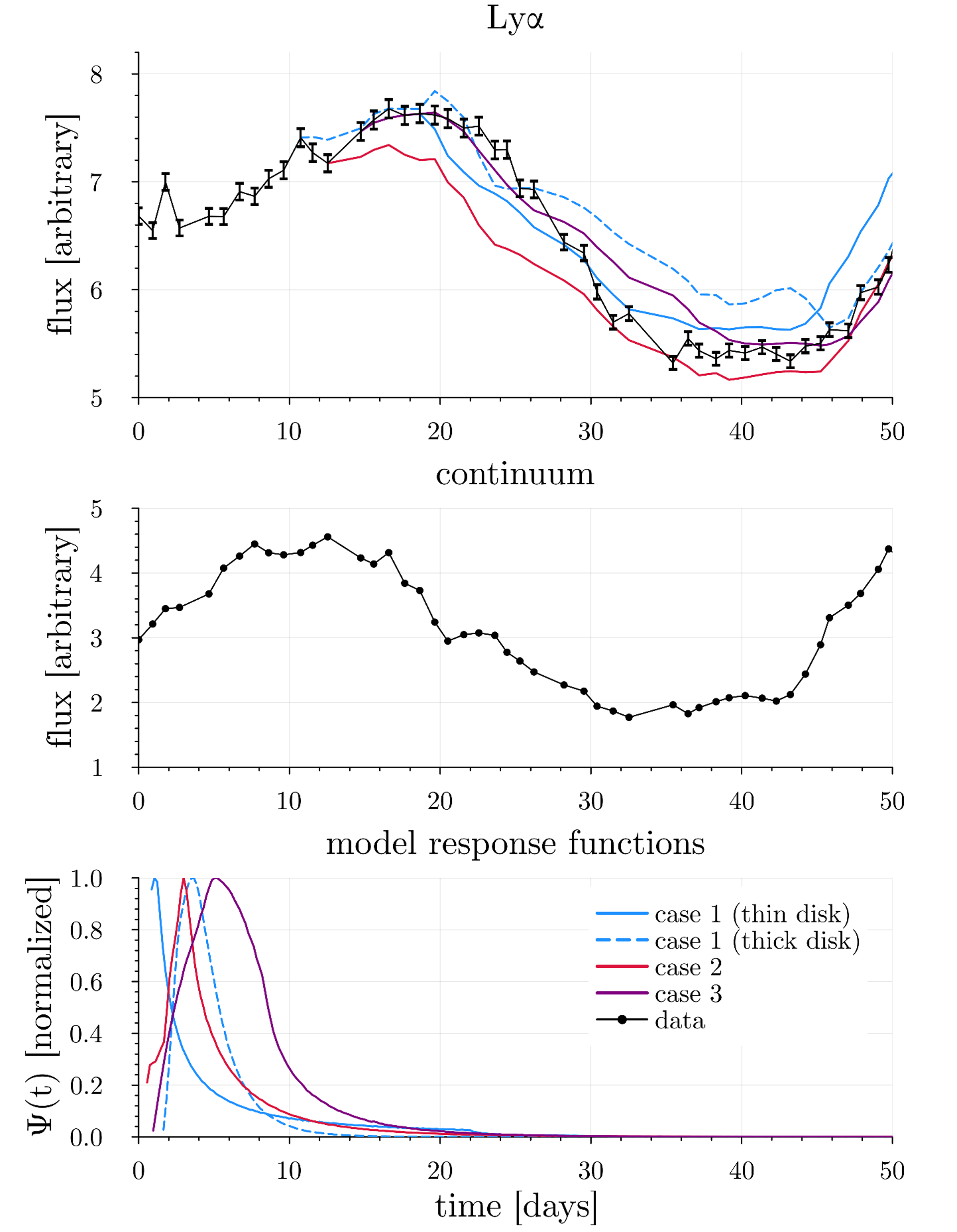}
        \caption{Sample synthetic lightcurves for each class of model. The time span is chosen to match the previous fits shown in \cite{STORM_CLOUDS}. Note that the \hyperref[case2]{case 2} and \hyperref[case3]{case 3} models appear to perform the best, and appear to be of similar quality to previous integrated synthetic lightcurves.}
        \label{fig:LC}
        \end{minipage}
    \end{figure}
    In order for the shorter delay component to be narrow in velocity space, the emitting gas must be in an area near the line of sight, i.e. near $\phi \approx 0$ in betwen the black hole and the observer. If the gas is restricted to this region it will have a small velocity spread along the line of sight regardless of distance from the black hole, allowing for a narrow width component with much shorter delays. One could imagine even more contrived versions of this kind of scenario that may allow the narrower velocity component to have shorter delays, but we do not consider them here. Just as in \hyperref[case2]{case 2}, we do not attempt to actually fit the data with this kind of scheme as if each cloud model is allowed to vary independently of the other the number of parameters increases quickly\textemdash just two cloud models can easily have more than 25 independent parameters. 
    
    Sample synthetic lightcurves for each of our model cases described above are shown in Figure \ref{fig:LC}. While none of the lightcurves match perfectly, \hyperref[case2]{case 2} and \hyperref[case3]{3} clearly perform the best. In comparing with previous work they appear to match the 1D RM data roughly as well as previously fit single-component cloud models of the BLR do, but they do so while maintaining a structure closer to what is observed in the delay profile. 

\section{Discussion and Implications}
Previous analyses of these data have attempted to recover model-agnostic 2D velocity-delay maps \citep{STORM_ECHO} and find best-fitting cloud models to the data \citep{STORM_CLOUDS}. Here we show that there is disagreement between modeling two popular models of the BLR (cloud or disk-wind) and the 2D data products, and discuss several possible geometric and kinematic BLR models that may alleviate this tension. Cases \hyperref[case2]{2} and \hyperref[case3]{3} are somewhat similar to two-component models of the BLR suggested in the past based on observations of the line profile alone \citep{2Component_Nagoshi2024_CSQ,2Component_Zhang_2011_ILR,2Component_NLR_Ludwig_Greene,2Component_Zhu_2009_ILR_VBLR,2Component_Brotherton1994_original}, but note that the cloud model can also satisfactorily explain the features observed in AGN broad-line profiles \citep{PANCOAST_CLOUDS_14,STORM_CLOUDS,2Component_GRAVITY}.

\textbf{This is why we focus on line \textit{and} delay profiles for various models\textemdash to fully understand the BLR and correctly model it, we must account for variability, thus the delay profile presents a powerful opportunity to better constrain the true nature of the BLR.} 
As the Lyman alpha delay profile in NGC 5548 shows two peaks in the data, this is strong evidence in favor of a multicomponent BLR in NGC 5548. We do not show 2D transfer functions for simplicity, as the delay profile alone demonstrates the problem with current modeling efforts succinctly and elegantly, but of course any model that correctly fits this profile would necessarily have a more accurate 2D transfer function as well. We also note that while
in this work we have focused mostly on the structural
information that can be learned about the BLR from the
combination of the line and delay profiles, this work also
demonstrates that it is critical for future modeling efforts
attempt to fully consider the variability of the BLR. This is in
part addressed through fitting the delay profile, but more
formally future modeling efforts should consider the RMS line
profile as well, which we omit here for the sake of simplicity
but again note that of course any model correctly matching
both the line and delay profiles will necessarily be a step in the
right direction toward this goal.

Note that there are likely other physical pictures that could be invoked to explain the discrepancy between the delay and line profiles. There have been many possible models of the BLR proposed in the literature \citep{ChajetHall2013,Flohic2012,WATERS16,PANCOAST_CLOUDS_14,2Component_Brotherton1994_original,CM96} and we have not tested all of them in this work, focusing solely on two popular classes of model that are numerically easy to employ. While we believe the discussion here to be robust and broadly applicable to cloud or disk-wind type models, even within these two models we have not explored every possible choice of parameters. We have also assumed here that the gas that responds is strongly correlated with the gas that creates the emission line profile. Perhaps this is not entirely true, or at least the relationship between the emitting gas and responding gas does not have a simple power-law radial dependence as assumed in this work. It may be possible that the line and delay profiles are not representative of the same gas populations, but one must invoke some complications to explain how this can be, and how it can be in such a manner that one produces a double-peak in the delay profile but a single peak in the line profile.

However, even if the responding gas and line-emitting gas are different populations, there is still an interesting problem posed above in regards to the measurement of the SMBH since the widths of the profiles are different from simple expectations given the same black hole mass. This picture likely also requires more complicated modeling to understand, but we do not explore it in this paper as we still believe it is more likely that the responding gas is significantly related to the line-emitting gas. 

While we have focused our discussion  on modeling the data obtained by the AGN STORM campaign on NGC 5548 \citep{STORM_MODEL_SPECTRA}, in particular on the Lyman $\alpha$ line as the morphology is simplest, the general trends observed in NGC 5548 are not unique. The C~{\sc iv} profile appears somewhat more complicated but similar in morphology, and while velocity resolved RM measurements are difficult and limited to a few objects, similar shapes in the delay and line profiles have been observed in 3C 273 and Mrk 817, for example, thus whatever is happening in NGC 5548 is likely more than just a one-off observation \citep{STORM2_II,RM3c27319}. 
    
If \textbf{\hyperref[case1]{case 1}} (large obscurer) were the correct prescription for the BLR, small changes in the obscuring medium would lead to large changes in the delay profile while leaving the line profile shape relatively unchanged. Physically this picture could be motivated by the presence of an obscurer\textemdash potentially a disk-wind being blown off of the BLR of which there is observational evidence for\textemdash that makes the back half of the system harder to see. A simple observational test of this model would be to watch for the variations in the measured delay profile as a function of velocity\textemdash with much larger variations expected near $v=0$ than near the peaks. In the limit where the obscurer were to evaporate the previous peaks in the delay profile should become the ``shoulders" of a newly single-peaked delay profile. This assumes that the obscuring medium is at least somewhat variable, but this seems like a reasonable assumption given the changes in luminosity, delays, and line-widths were observed over months-long timescales in the BLR. In the data for NGC 5548, the C~{\sc iv} delay profile shows significant variability across the observations, notably near line center as predicted above, yet the Lyman $\alpha$ profile morphology does not change much, making it more likely that the C~{\sc iv} profile is contaminated by inflows and outflows. 

Similarly for  \textbf{\hyperref[case3]{case 3}}, any changes in the properties of the concentrated component of would dramatically impact observations near line-center. In our \hyperref[case3]{case 3} model we propose that the responding gas in the narrow velocity component be constrained to a region close to the line of sight between the black hole and the observer, which allows for the observed velocity spread to be small and the delays short. This is difficult to imagine a physical mechanism that would concentrate the emitting gas in the narrow velocity component in an area of low delay space so close to the line of sight in between the black hole and the observer. While the easiest way to produce a narrow component in velocity space is simply to put the emitting gas far away, if one does this the delays of the narrow width component will naturally be larger than the broader component, the opposite of the desired behavior. 

If \textbf{\hyperref[case2]{case 2}} (thin disk + clouds) were the correct prescription for the BLR, we would expect to observe more static delay and line profiles, with changes in delay amplitude corresponding to the movement of the characteristic size of the BLR components. The Lyman $\alpha$ delay profile showing two peaks across the entire observation period is consistent with this interpretation, but as noted above the picture presented in C~{\sc iv} \citep{STORM_MODEL_SPECTRA} is messier. The inflow component in this model is also in keeping with the physical picture of the two component model\textemdash inflows are necessary to feed the central accretion disk, thus the downward slope in the delay profile could represent a part of the story of how gas moves from the torus to the inner accretion disk.

When comparing the locations of the peaks in the observed delay profile to the corresponding locations in the observed line profile, we see that interestingly they occur near a flux value corresponding to $\sim$ 50\% of the observed strength\textemdash thus there must be a large component of gas emitting relatively close to the SMBH to produce the dip in the delay profile between the peaks. The modeled line profile variability in STORM VIII \citep{STORM_MODEL_SPECTRA} also supports this behavior\textemdash the variability peaks near $v=0$, where large quantities of gas emit at relatively short delays, lending further support to either our \hyperref[case2]{case 2} or \hyperref[case3]{case 3} interpretations of the BLR presented here.

Such detailed modeling of the BLR is critical to understand its true nature, but may be less important in constraining the black hole mass. While the three cases described in this work are just schematic pictures and we cannot entirely rule out tension in black hole mass measurements without properly fitting them to the data, in our simple models our black hole mass changes by only a factor of a few, at most, from that presented in previous work. For example, to obtain the delay profiles shown in Figure \ref{fig:case2} requires a black hole mass $\sim$2-3$\times$ larger than previously published. While these more complicated models may better explain the fundamental nature of the BLR, they introduce at most a factor of a few tension in measuring the masses SMBHs. Thus to obtain the most precise mass estimates and understand this systematic error we must understand the true nature of the BLR through cohesive models.

Finally, the reason why the simple disk-wind model used in \cite{Long_2023} was able to successfully fit GRAVITY BLR data but not all of the RM data products is also interesting: in marginally resolved first-order interferometry we are most sensitive to the sides of the system where there is still flux but the wavelength shift is large. Conversely such measurements are not as sensitive to the front/back of the system at line center. RM is only restricted by the sampling and total duration for the time baseline, but these restrictions apply equally at line center as they do in the wings. Thus RM measurements\textemdash in particular near line-center\textemdash complement current marginally resolved GRAVITY results and can help rule out competing interpretations of the true geometry and kinematics of the BLR. In the future, advances in observational capabilities in GRAVITY or other interferometric instruments that can resolve second or higher moments of the BLR image will also be able to better constrain which classes of models best represent the true nature of the BLR. This is important beyond simply measuring the masses of black holes, as the combination of GRAVITY and RM measurements also allow for an interesting new test of the Hubble constant \citep{SARM_2025}. Such measurements require a model of the BLR that can map from RM measured sizes to GRAVITY measured sizes and may contain unexplored systematics \citep{Li_2025_Response}. 

At the highest level, the analysis in this paper demonstrates that the delay and line profiles in NGC 5548 imply the BLR is likely not a single distribution of puffed up clouds nor a simple disk-wind type model, but instead a more complicated combination of the two or perhaps something even more exotic not considered here. Future observations and modeling efforts are needed to constrain which picture best represents the fundamental geometry and kinematics of the BLR, as well as to better constrain the systematic errors introduced through our choice of model.

\section{Acknowledgments}
This work was supported in part by NSF grants AST-1909711 and AST-2307983, and by an Alfred P. Sloan Research Fellowship (JD). All the {\it HST} data used in this work can be found on the MAST archive:  \dataset[10.17909/t9-ky1s-j932]{http://dx.doi.org/10.17909/t9-ky1s-j932} This work utilized the Alpine high performance computing resource at the University of Colorado Boulder. Alpine is jointly funded by the University of Colorado Boulder, the University of Colorado Anschutz, and Colorado State University and with support from NSF grants OAC-2201538 and OAC-2322260. KL is especially grateful to Sajal Gupta, Gisella de Rosa, Chris Done, and Jian-Min Wang for many helpful discussions over the course of the project. 

We are grateful to the anonymous referee for their thorough report, which greatly improved the quality of the paper. The code used in this work is available free and open-source on \href{https://github.com/kirklong/BroadLineRegions.jl}{GitHub}. A plain-language summary of this work can be found at \href{www.kirklong.space/RMProblem.html}{on the author's website}. 

\software{Julia,
          \texttt{BroadLineRegions.jl}
          }

\newpage

\bibliography{citations}{}

\begin{thebibliography}{}
\expandafter\ifx\csname natexlab\endcsname\relax\def\natexlab#1{#1}\fi
\providecommand{\url}[1]{\href{#1}{#1}}
\providecommand{\dodoi}[1]{doi:~\href{http://doi.org/#1}{\nolinkurl{#1}}}
\providecommand{\doeprint}[1]{\href{http://ascl.net/#1}{\nolinkurl{http://ascl.net/#1}}}
\providecommand{\doarXiv}[1]{\href{https://arxiv.org/abs/#1}{\nolinkurl{https://arxiv.org/abs/#1}}}

\bibitem[{G. {Batchelor}(1968){Batchelor}}]{Batchelor68}
{Batchelor}, G. 1968, \bibinfo{title}{{An introduction to fluid dynamics. G. K. Batchelor, F.R.S., London (Cambridge University Press), 1967. Pp. xvii, 615; Plates 24; Numerous Figures. 75s. in U.K., \$13.50 in U.S.A},} Quarterly Journal of the Royal Meteorological Society, 94, 435, \dodoi{10.1002/qj.49709440128}

\bibitem[{M.~C. {Bentz} {et~al.}(2013){Bentz}, {Denney}, {Grier}, {Barth}, {Peterson}, {Vestergaard}, {Bennert}, {Canalizo}, {De Rosa}, {Filippenko}, {Gates}, {Greene}, {Li}, {Malkan}, {Pogge}, {Stern}, {Treu}, \& {Woo}}]{Bentz2013_Hbeta_R-L}
{Bentz}, M.~C., {Denney}, K.~D., {Grier}, C.~J., {et~al.} 2013, \bibinfo{title}{{The Low-luminosity End of the Radius-Luminosity Relationship for Active Galactic Nuclei},} \apj, 767, 149, \dodoi{10.1088/0004-637X/767/2/149}

\bibitem[{R.~D. {Blandford} \& C.~F. {McKee}(1982){Blandford} \& {McKee}}]{BLANDFORD_MCKEE_82}
{Blandford}, R.~D., \& {McKee}, C.~F. 1982, \bibinfo{title}{{Reverberation mapping of the emission line regions of Seyfert galaxies and quasars.},} \apj, 255, 419, \dodoi{10.1086/159843}

\bibitem[{M.~S. {Brotherton} {et~al.}(1994){Brotherton}, {Wills}, {Francis}, \& {Steidel}}]{2Component_Brotherton1994_original}
{Brotherton}, M.~S., {Wills}, B.~J., {Francis}, P.~J., \& {Steidel}, C.~C. 1994, \bibinfo{title}{{The Intermediate Line Region of QSOs},} \apj, 430, 495, \dodoi{10.1086/174425}

\bibitem[{E.~M. {Cackett} {et~al.}(2021){Cackett}, {Bentz}, \& {Kara}}]{RM_Review_21}
{Cackett}, E.~M., {Bentz}, M.~C., \& {Kara}, E. 2021, \bibinfo{title}{{Reverberation mapping of active galactic nuclei: from X-ray corona to dusty torus},} iScience, 24, 102557, \dodoi{10.1016/j.isci.2021.102557}

\bibitem[{L.~S. {Chajet} \& P.~B. {Hall}(2013){Chajet} \& {Hall}}]{ChajetHall2013}
{Chajet}, L.~S., \& {Hall}, P.~B. 2013, \bibinfo{title}{{Magnetohydrodynamic disc winds and linewidth distributions},} \mnras, 429, 3214, \dodoi{10.1093/mnras/sts580}

\bibitem[{J. {Chiang} \& N. {Murray}(1996){Chiang} \& {Murray}}]{CM96}
{Chiang}, J., \& {Murray}, N. 1996, \bibinfo{title}{{Reverberation Mapping and the Disk-Wind Model of the Broad-Line Region},} \apj, 466, 704, \dodoi{10.1086/177543}

\bibitem[{H. {Cho} {et~al.}(2023){Cho}, {Woo}, {Wang}, {Son}, {Shin}, {Rakshit}, {Barth}, {Bennert}, {Gallo}, {Hodges-Kluck}, {Treu}, {Bae}, {Cho}, {Foord}, {Geum}, {Jadhav}, {Jeon}, {Kabasares}, {Kang}, {Kang}, {Kim}, {Kim}, {Kim}, {Kim}, {N. Le}, {Malkan}, {Mandal}, {Park}, {Park}, {Sung}, {U}, \& {Williams}}]{Halpha_R-L_Cho_2023}
{Cho}, H., {Woo}, J.-H., {Wang}, S., {et~al.} 2023, \bibinfo{title}{{The Seoul National University AGN Monitoring Project. IV. H{\ensuremath{\alpha}} Reverberation Mapping of Six AGNs and the H{\ensuremath{\alpha}} Size-Luminosity Relation},} \apj, 953, 142, \dodoi{10.3847/1538-4357/ace1e5}

\bibitem[{ {Dalla Bontà, E.} {et~al.}(2025){Dalla Bontà, E.}, {Peterson, B. M.}, {Grier, C. J.}, {Berton, M.}, {Brandt, W. N.}, {Ciroi, S.}, {Corsini, E. M.}, {Dalla Barba, B.}, {Davies, R.}, {Dehghanian, M.}, {Edelson, R.}, {Foschini, L.}, {Gasparri, D.}, {Ho, L. C.}, {Horne, K.}, {Iodice, E.}, {Morelli, L.}, {Pizzella, A.}, {Portaluri, E.}, {Shen, Y.}, {Schneider, D. P.}, \& {Vestergaard, M.}}]{2024arXivDallaBonta}
{Dalla Bontà, E.}, {Peterson, B. M.}, {Grier, C. J.}, {et~al.} 2025, \bibinfo{title}{{Estimating masses of supermassive black holes in active galactic nuclei from the Hα emission line},} A\&A, 696, A48, \dodoi{10.1051/0004-6361/202452746}

\bibitem[{G. {De Rosa} {et~al.}(2015){De Rosa}, {Peterson}, {Ely}, {Kriss}, {Crenshaw}, {Horne}, {Korista}, {Netzer}, {Pogge}, {Ar{\'e}valo}, {Barth}, {Bentz}, {Brandt}, {Breeveld}, {Brewer}, {Dalla Bont{\`a}}, {De Lorenzo-C{\'a}ceres}, {Denney}, {Dietrich}, {Edelson}, {Evans}, {Fausnaugh}, {Gehrels}, {Gelbord}, {Goad}, {Grier}, {Grupe}, {Hall}, {Kaastra}, {Kelly}, {Kennea}, {Kochanek}, {Lira}, {Mathur}, {McHardy}, {Nousek}, {Pancoast}, {Papadakis}, {Pei}, {Schimoia}, {Siegel}, {Starkey}, {Treu}, {Uttley}, {Vaughan}, {Vestergaard}, {Villforth}, {Yan}, {Young}, \& {Zu}}]{STORM_CCF}
{De Rosa}, G., {Peterson}, B.~M., {Ely}, J., {et~al.} 2015, \bibinfo{title}{{Space Telescope and Optical Reverberation Mapping Project.I. Ultraviolet Observations of the Seyfert 1 Galaxy NGC 5548 with the Cosmic Origins Spectrograph on Hubble Space Telescope},} \apj, 806, 128, \dodoi{10.1088/0004-637X/806/1/128}

\bibitem[{H.~M.~L.~G. {Flohic} {et~al.}(2012){Flohic}, {Eracleous}, \& {Bogdanovi{\'c}}}]{Flohic2012}
{Flohic}, H. M.~L.~G., {Eracleous}, M., \& {Bogdanovi{\'c}}, T. 2012, \bibinfo{title}{{Effects of an Accretion Disk Wind on the Profile of the Balmer Emission Lines from Active Galactic Nuclei},} \apj, 753, 133, \dodoi{10.1088/0004-637X/753/2/133}

\bibitem[{M.~R. Goad \& K.~T. Korista(2014)Goad \& Korista}]{Goad_Korista_response_2014}
Goad, M.~R., \& Korista, K.~T. 2014, \bibinfo{title}{Interpreting broad emission-line variations – I. Factors influencing the emission-line response,} Monthly Notices of the Royal Astronomical Society, 444, 43, \dodoi{10.1093/mnras/stu1456}

\bibitem[{ {GRAVITY Collaboration} {et~al.}(2017){GRAVITY Collaboration}, {Abuter}, {Accardo}, {Amorim}, {Anugu}, {{\'A}vila}, {Azouaoui}, {Benisty}, {Berger}, {Blind}, {Bonnet}, {Bourget}, {Brandner}, {Brast}, {Buron}, {Burtscher}, {Cassaing}, {Chapron}, {Choquet}, {Cl{\'e}net}, {Collin}, {Coud{\'e} Du Foresto}, {de Wit}, {de Zeeuw}, {Deen}, {Delplancke-Str{\"o}bele}, {Dembet}, {Derie}, {Dexter}, {Duvert}, {Ebert}, {Eckart}, {Eisenhauer}, {Esselborn}, {F{\'e}dou}, {Finger}, {Garcia}, {Garcia Dabo}, {Garcia Lopez}, {Gendron}, {Genzel}, {Gillessen}, {Gonte}, {Gordo}, {Grould}, {Gr{\"o}zinger}, {Guieu}, {Haguenauer}, {Hans}, {Haubois}, {Haug}, {Haussmann}, {Henning}, {Hippler}, {Horrobin}, {Huber}, {Hubert}, {Hubin}, {Hummel}, {Jakob}, {Janssen}, {Jochum}, {Jocou}, {Kaufer}, {Kellner}, {Kendrew}, {Kern}, {Kervella}, {Kiekebusch}, {Klein}, {Kok}, {Kolb}, {Kulas}, {Lacour}, {Lapeyr{\`e}re}, {Lazareff}, {Le Bouquin}, {L{\`e}na}, {Lenzen}, {L{\'e}v{\^e}que}, {Lippa}, {Magnard}, {Mehrgan}, {Mellein}, {M{\'e}rand},
  {Moreno-Ventas}, {Moulin}, {M{\"u}ller}, {M{\"u}ller}, {Neumann}, {Oberti}, {Ott}, {Pallanca}, {Panduro}, {Pasquini}, {Paumard}, {Percheron}, {Perraut}, {Perrin}, {Pfl{\"u}ger}, {Pfuhl}, {Phan Duc}, {Plewa}, {Popovic}, {Rabien}, {Ram{\'\i}rez}, {Ramos}, {Rau}, {Riquelme}, {Rohloff}, {Rousset}, {Sanchez-Bermudez}, {Scheithauer}, {Sch{\"o}ller}, {Schuhler}, {Spyromilio}, {Straubmeier}, {Sturm}, {Suarez}, {Tristram}, {Ventura}, {Vincent}, {Waisberg}, {Wank}, {Weber}, {Wieprecht}, {Wiest}, {Wiezorrek}, {Wittkowski}, {Woillez}, {Wolff}, {Yazici}, {Ziegler}, \& {Zins}}]{GRAVITY17}
{GRAVITY Collaboration}, {Abuter}, R., {Accardo}, M., {et~al.} 2017, \bibinfo{title}{{First light for GRAVITY: Phase referencing optical interferometry for the Very Large Telescope Interferometer},} \aap, 602, A94, \dodoi{10.1051/0004-6361/201730838}

\bibitem[{ {GRAVITY Collaboration} {et~al.}(2019){GRAVITY Collaboration}, {Abuter}, {Accardo}, {Adler}, {Amorim}, {Anugu}, {{\'A}vila}, {Baub{\"o}ck}, {Benisty}, {Berger}, {Bestenlehner}, {Beust}, {Blind}, {Bonnefoy}, {Bonnet}, {Bourget}, {Bouvier}, {Brandner}, {Brast}, {Buron}, {Burtscher}, {Cantalloube}, {Caratti O Garatti}, {Caselli}, {Cassaing}, {Chapron}, {Charnay}, {Choquet}, {Cl{\'e}net}, {Collin}, {Coud{\'e} Du Foresto}, {Davies}, {Deen}, {Delplancke-Str{\"o}bele}, {Dembet}, {Derie}, {de Wit}, {Dexter}, {de Zeeuw}, {Dougados}, {Dubus}, {Duvert}, {Ebert}, {Eckart}, {Eisenhauer}, {Esselborn}, {Eupen}, {F{\'e}dou}, {Ferreira}, {Finger}, {F{\"o}rster Schreiber}, {Gao}, {Garc{\'\i}a Dab{\'o}}, {Garcia Lopez}, {Garcia}, {Gendron}, {Genzel}, {Gerhard}, {Gil}, {Gillessen}, {Gont{\'e}}, {Gordo}, {Gratadour}, {Greenbaum}, {Grellmann}, {Gr{\"o}zinger}, {Guajardo}, {Guieu}, {Habibi}, {Haguenauer}, {Hans}, {Haubois}, {Haug}, {Hau{\ss}mann}, {Henning}, {Hippler}, {H{\"o}nig}, {Horrobin}, {Huber}, {Hubert}, {Hubin},
  {Hummel}, {Jakob}, {Janssen}, {Jimenez Rosales}, {Jochum}, {Jocou}, {Kammerer}, {Karl}, {Kaufer}, {Kellner}, {Kendrew}, {Kern}, {Kervella}, {Kiekebusch}, {Kishimoto}, {Klarmann}, {Klein}, {K{\"o}hler}, {Kok}, {Kolb}, {Koutoulaki}, {Kulas}, {Labadie}, {Lacour}, {Lagrange}, {Lapeyr{\`e}re}, {Laun}, {Lazareff}, {Le Bouquin}, {L{\'e}na}, {Lenzen}, {L{\'e}v{\^e}que}, {Lin}, {Lippa}, {Lutz}, {Magnard}, {Maire}, {Mehrgan}, {M{\'e}rand}, {Millour}, {Molli{\`e}re}, {Moulin}, {M{\"u}ller}, {M{\"u}ller}, {M{\"u}ller}, {Netzer}, {Neumann}, {Nowak}, {Oberti}, {Ott}, {Pallanca}, {Panduro}, {Pasquini}, {Paumard}, {Percheron}, {Perraut}, {Perrin}, {Peterson}, {Petrucci}, {Pfl{\"u}ger}, {Pfuhl}, {Phan Duc}, {Pineda}, {Plewa}, {Popovic}, {Pott}, {Prieto}, {Pueyo}, {Rabien}, {Ram{\'\i}rez}, {Ramos}, {Rau}, {Ray}, {Riquelme}, {Rodr{\'\i}guez-Coira}, {Rohloff}, {Rouan}, {Rousset}, {Sanchez-Bermudez}, {Schartmann}, {Scheithauer}, {Sch{\"o}ller}, {Schuhler}, {Segura-Cox}, {Shangguan}, {Shimizu}, {Spyromilio}, {Sternberg},
  {Stock}, {Straub}, {Straubmeier}, {Sturm}, {Su{\'a}rez Valles}, {Tacconi}, {Thi}, {Tristram}, {Valenzuela}, {van Boekel}, {van Dishoeck}, {Vermot}, {Vincent}, {von Fellenberg}, {Waisberg}, {Wang}, {Wank}, {Weber}, {Weigelt}, {Widmann}, {Wieprecht}, {Wiest}, {Wiezorrek}, {Wittkowski}, {Woillez}, {Wolff}, {Yang}, {Yazici}, {Ziegler}, \& {Zins}}]{GRAVITY19}
{GRAVITY Collaboration}, {Abuter}, R., {Accardo}, M., {et~al.} 2019, \bibinfo{title}{{Spatially Resolving the Quasar Broad Emission Line Region},} The Messenger, 178, 20, \dodoi{10.18727/0722-6691/5166}

\bibitem[{ {GRAVITY+ Collaboration} {et~al.}(2023){GRAVITY+ Collaboration}, {:}, {Abuter}, {Alarcon}, {Allouche}, {Amorim}, {Bailet}, {Bedigan}, {Berdeu}, {Berger}, {Berio}, {Bigioli}, {Blaho}, {Boebion}, {Bolzer}, {Bonnet}, {Bourdarot}, {Bourget}, {Brandner}, {Cardenas}, {Conzelmann}, {Comin}, {Cl{\'e}net}, {Courtney-Barrer}, {Dallilar}, {Davies}, {Defr{\`e}re}, {Delboulb{\'e}}, {Delplancke-Str{\"o}bele}, {Dembet}, {de Zeeuw}, {Drescher}, {Eckart}, {{\'E}douard}, {Eisenhauer}, {Fabricius}, {Feuchtgruber}, {Finger}, {F{\"o}rster Schreiber}, {Fuenteseca}, {Garcia}, {Garcia}, {Gao}, {Gendron}, {Genzel}, {Gil}, {Gillessen}, {Gomes}, {Gont{\'e}}, {Gouvret}, {Guajardo}, {Guidolin}, {Guieu}, {Guzmann}, {Hackenberg}, {Haddad}, {Hartl}, {Haubois}, {Hau{\ss}mann}, {Hei{\ss}el}, {Henning}, {Hippler}, {H{\"o}nig}, {Horrobin}, {Hubin}, {Jacqmart}, {Jocou}, {Kaufer}, {Kervella}, {Kirchbauer}, {Kolb}, {Korhonen}, {Kreidberg}, {Krempl}, {Lacour}, {Lagarde}, {Lai}, {Lapeyr{\`e}re}, {Laugier}, {Le Bouquin}, {Leftley},
  {L{\'e}na}, {Lewis}, {Lutz}, {Magnard}, {Mang}, {Marcotto}, {Maurel}, {M{\'e}rand}, {Millour}, {More}, {Nowack}, {Nowak}, {Oberti}, {Olivares}, {Ott}, {Pallanca}, {Paumard}, {Perraut}, {Perrin}, {Petrov}, {Pfuhl}, {Pourr{\'e}}, {Rabien}, {Rau}, {Riquelme}, {Robbe-Dubois}, {Rochat}, {Salman}, {Scherbarth}, {Sch{\"o}ller}, {Schubert}, {Schuhler}, {Shangguan}, {Shchekaturov}, {Shimizu}, {Scheithauer}, {Sevin}, {Soenke}, {Soulez}, {Spang}, {Stadler}, {Straubmeier}, {Sturm}, {Sykes}, {Tacconi}, {Tischer}, {Tristram}, {Vincent}, {von Fellenberg}, {Uysal}, {Widmann}, {Wieprecht}, {Wiezorrek}, {Woillez}, {Yaz{\i}c{\i}}, \& {Zins}}]{GRAVITY+}
{GRAVITY+ Collaboration}, {:}, {Abuter}, R., {et~al.} 2023, \bibinfo{title}{{The GRAVITY+ Project: Towards All-sky, Faint-Science, High-Contrast Near-Infrared Interferometry at the VLTI},} arXiv e-prints, arXiv:2301.08071, \dodoi{10.48550/arXiv.2301.08071}

\bibitem[{ {GRAVITY Collaboration} {et~al.}(2024){GRAVITY Collaboration}, {Amorim, A.}, {Bourdarot, G.}, {Brandner, W.}, {Cao, Y.}, {Clénet, Y.}, {Davies, R.}, {de Zeeuw, P. T.}, {Dexter, J.}, {Drescher, A.}, {Eckart, A.}, {Eisenhauer, F.}, {Fabricius, M.}, {Feuchtgruber, H.}, {Förster Schreiber, N. M.}, {Garcia, P. J. V.}, {Genzel, R.}, {Gillessen, S.}, {Gratadour, D.}, {Hönig, S.}, {Kishimoto, M.}, {Lacour, S.}, {Lutz, D.}, {Millour, F.}, {Netzer, H.}, {Ott, T.}, {Paumard, T.}, {Perraut, K.}, {Perrin, G.}, {Peterson, B. M.}, {Petrucci, P. O.}, {Pfuhl, O.}, {Prieto, M. A.}, {Rabien, S.}, {Rouan, D.}, {Santos, D. J. D.}, {Shangguan, J.}, {Shimizu, T.}, {Sternberg, A.}, {Straubmeier, C.}, {Sturm, E.}, {Tacconi, L. J.}, {Tristram, K. R. W.}, {Widmann, F.}, \& {Woillez, J.}}]{GRAVITY_R-L_2024}
{GRAVITY Collaboration}, {Amorim, A.}, {Bourdarot, G.}, {et~al.} 2024, \bibinfo{title}{The size-luminosity relation of local active galactic nuclei from interferometric observations of the broad-line region⋆,} A\&A, 684, A167, \dodoi{10.1051/0004-6361/202348167}

\bibitem[{C.~J. Grier {et~al.}(2013)Grier, Peterson, Horne, Bentz, Pogge, Denney, De~Rosa, Martini, Kochanek, Zu, Shappee, Siverd, Beatty, Sergeev, Kaspi, Salvo, Bird, Bord, Borman, Che, Chen, Cohen, Dietrich, Doroshenko, Efimov, Free, Ginsburg, Henderson, King, Mogren, Molina, Mosquera, Nazarov, Okhmat, Pejcha, Rafter, Shields, Skowron, Szczygiel, Valluri, \& van Saders}]{Grier_2013_ECHO_INFALL}
Grier, C.~J., Peterson, B.~M., Horne, K., {et~al.} 2013, \bibinfo{title}{THE STRUCTURE OF THE BROAD-LINE REGION IN ACTIVE GALACTIC NUCLEI. I. RECONSTRUCTED VELOCITY-DELAY MAPS,} The Astrophysical Journal, 764, 47, \dodoi{10.1088/0004-637X/764/1/47}

\bibitem[{Y. {Homayouni} {et~al.}(2023){Homayouni}, {De Rosa}, {Plesha}, {Kriss}, {Barth}, {Cackett}, {Horne}, {Kara}, {Landt}, {Arav}, {Boizelle}, {Bentz}, {Brink}, {Brotherton}, {Chelouche}, {Dalla Bont{\`a}}, {Dehghanian}, {Du}, {Ferland}, {Ferrarese}, {Fian}, {Filippenko}, {Fischer}, {Foley}, {Gelbord}, {Goad}, {Gonz{\'a}lez Buitrago}, {Gorjian}, {Grier}, {Hall}, {Hern{\'a}ndez Santisteban}, {Hu}, {Ili{\'c}}, {Joner}, {Kaastra}, {Kaspi}, {Kochanek}, {Korista}, {Kova{\v{c}}evi{\'c}}, {Kynoch}, {Li}, {McHardy}, {McLane}, {Mehdipour}, {Miller}, {Mitchell}, {Montano}, {Netzer}, {Panagiotou}, {Partington}, {Pogge}, {{\v{C}}. Popovi{\'c}}, {Proga}, {Rogantini}, {Storchi-Bergmann}, {Sanmartim}, {Siebert}, {Treu}, {Vestergaard}, {Wang}, {Ward}, {Waters}, {Williams}, {Zaidouni}, \& {Zu}}]{STORM2_II}
{Homayouni}, Y., {De Rosa}, G., {Plesha}, R., {et~al.} 2023, \bibinfo{title}{{AGN STORM 2. II. Ultraviolet Observations of Mrk 817 with the Cosmic Origins Spectrograph on the Hubble Space Telescope},} \apj, 948, 85, \dodoi{10.3847/1538-4357/acc45a}

\bibitem[{K. {Horne}(1994){Horne}}]{memecho2}
{Horne}, K. 1994, in Astronomical Society of the Pacific Conference Series, Vol.~69, Reverberation Mapping of the Broad-Line Region in Active Galactic Nuclei, ed. P.~M. {Gondhalekar}, K.~{Horne}, \& B.~M. {Peterson}, 23

\bibitem[{K. {Horne} \& T.~R. {Marsh}(1986){Horne} \& {Marsh}}]{HorneMarsh86_KeplerianShear}
{Horne}, K., \& {Marsh}, T.~R. 1986, \bibinfo{title}{{Emission line formation in accretion discs},} \mnras, 218, 761, \dodoi{10.1093/mnras/218.4.761}

\bibitem[{K. {Horne} {et~al.}(2021){Horne}, {De Rosa}, {Peterson}, {Barth}, {Ely}, {Fausnaugh}, {Kriss}, {Pei}, {Bentz}, {Cackett}, {Edelson}, {Eracleous}, {Goad}, {Grier}, {Kaastra}, {Kochanek}, {Krongold}, {Mathur}, {Netzer}, {Proga}, {Tejos}, {Vestergaard}, {Villforth}, {Adams}, {Anderson}, {Ar{\'e}valo}, {Beatty}, {Bennert}, {Bigley}, {Bisogni}, {Borman}, {Boroson}, {Bottorff}, {Brandt}, {Breeveld}, {Brotherton}, {Brown}, {Brown}, {Canalizo}, {Carini}, {Clubb}, {Comerford}, {Corsini}, {Crenshaw}, {Croft}, {Croxall}, {Dalla Bont{\`a}}, {Deason}, {Dehghanian}, {De Lorenzo-C{\'a}ceres}, {Denney}, {Dietrich}, {Done}, {Efimova}, {Evans}, {Ferland}, {Filippenko}, {Flatland}, {Fox}, {Gardner}, {Gates}, {Gehrels}, {Geier}, {Gelbord}, {Gonzalez}, {Gorjian}, {Greene}, {Grupe}, {Gupta}, {Hall}, {Henderson}, {Hicks}, {Holmbeck}, {Holoien}, {Hutchison}, {Im}, {Jensen}, {Johnson}, {Joner}, {Jones}, {Kaspi}, {Kelly}, {Kennea}, {Kim}, {Kim}, {Kim}, {King}, {Klimanov}, {Korista}, {Lau}, {Lee}, {Leonard}, {Li}, {Lira},
  {Lochhaas}, {Ma}, {MacInnis}, {Malkan}, {Manne-Nicholas}, {Mauerhan}, {McGurk}, {McHardy}, {Montuori}, {Morelli}, {Mosquera}, {Mudd}, {M{\"u}ller-S{\'a}nchez}, {Nazarov}, {Norris}, {Nousek}, {Nguyen}, {Ochner}, {Okhmat}, {Pancoast}, {Papadakis}, {Parks}, {Penny}, {Pizzella}, {Pogge}, {Poleski}, {Pott}, {Rafter}, {Rix}, {Runnoe}, {Saylor}, {Schimoia}, {Schn{\"u}lle}, {Scott}, {Sergeev}, {Shappee}, {Shivvers}, {Siegel}, {Simonian}, {Siviero}, {Skielboe}, {Somers}, {Spencer}, {Starkey}, {Stevens}, {Sung}, {Tayar}, {Treu}, {Turner}, {Uttley}, {Van Saders}, {Vican}, {Villanueva}, {Weiss}, {Woo}, {Yan}, {Young}, {Yuk}, {Zheng}, {Zhu}, \& {Zu}}]{STORM_ECHO}
{Horne}, K., {De Rosa}, G., {Peterson}, B.~M., {et~al.} 2021, \bibinfo{title}{{Space Telescope and Optical Reverberation Mapping Project. IX. Velocity-Delay Maps for Broad Emission Lines in NGC 5548},} \apj, 907, 76, \dodoi{10.3847/1538-4357/abce60}

\bibitem[{K.~T. Korista \& M.~R. Goad(2004)Korista \& Goad}]{Korista_Goad_response_2004}
Korista, K.~T., \& Goad, M.~R. 2004, \bibinfo{title}{What the Optical Recombination Lines Can Tell Us about the Broad-Line Regions of Active Galactic Nuclei,} The Astrophysical Journal, 606, 749, \dodoi{10.1086/383193}

\bibitem[{G.~A. {Kriss} {et~al.}(2019){Kriss}, {De Rosa}, {Ely}, {Peterson}, {Kaastra}, {Mehdipour}, {Ferland}, {Dehghanian}, {Mathur}, {Edelson}, {Korista}, {Arav}, {Barth}, {Bentz}, {Brandt}, {Crenshaw}, {Dalla Bont{\`a}}, {Denney}, {Done}, {Eracleous}, {Fausnaugh}, {Gardner}, {Goad}, {Grier}, {Horne}, {Kochanek}, {McHardy}, {Netzer}, {Pancoast}, {Pei}, {Pogge}, {Proga}, {Silva}, {Tejos}, {Vestergaard}, {Adams}, {Anderson}, {Ar{\'e}valo}, {Beatty}, {Behar}, {Bennert}, {Bianchi}, {Bigley}, {Bisogni}, {Boissay-Malaquin}, {Borman}, {Bottorff}, {Breeveld}, {Brotherton}, {Brown}, {Brown}, {Cackett}, {Canalizo}, {Cappi}, {Carini}, {Clubb}, {Comerford}, {Coker}, {Corsini}, {Costantini}, {Croft}, {Croxall}, {Deason}, {De Lorenzo-C{\'a}ceres}, {De Marco}, {Dietrich}, {Di Gesu}, {Ebrero}, {Evans}, {Filippenko}, {Flatland}, {Gates}, {Gehrels}, {Geier}, {Gelbord}, {Gonzalez}, {Gorjian}, {Grupe}, {Gupta}, {Hall}, {Henderson}, {Hicks}, {Holmbeck}, {Holoien}, {Hutchison}, {Im}, {Jensen}, {Johnson}, {Joner}, {Kaspi},
  {Kelly}, {Kelly}, {Kennea}, {Kim}, {Kim}, {Kim}, {King}, {Klimanov}, {Krongold}, {Lau}, {Lee}, {Leonard}, {Li}, {Lira}, {Lochhaas}, {Ma}, {MacInnis}, {Malkan}, {Manne-Nicholas}, {Matt}, {Mauerhan}, {McGurk}, {Montuori}, {Morelli}, {Mosquera}, {Mudd}, {M{\"u}ller-S{\'a}nchez}, {Nazarov}, {Norris}, {Nousek}, {Nguyen}, {Ochner}, {Okhmat}, {Paltani}, {Parks}, {Pinto}, {Pizzella}, {Poleski}, {Ponti}, {Pott}, {Rafter}, {Rix}, {Runnoe}, {Saylor}, {Schimoia}, {Schn{\"u}lle}, {Scott}, {Sergeev}, {Shappee}, {Shivvers}, {Siegel}, {Simonian}, {Siviero}, {Skielboe}, {Somers}, {Spencer}, {Starkey}, {Stevens}, {Sung}, {Tayar}, {Teems}, {Treu}, {Turner}, {Uttley}, {. Van Saders}, {Vican}, {Villforth}, {Villanueva}, {Walton}, {Waters}, {Weiss}, {Woo}, {Yan}, {Yuk}, {Zheng}, {Zhu}, \& {Zu}}]{STORM_MODEL_SPECTRA}
{Kriss}, G.~A., {De Rosa}, G., {Ely}, J., {et~al.} 2019, \bibinfo{title}{{Space Telescope and Optical Reverberation Mapping Project. VIII. Time Variability of Emission and Absorption in NGC 5548 Based on Modeling the Ultraviolet Spectrum},} \apj, 881, 153, \dodoi{10.3847/1538-4357/ab3049}

\bibitem[{Y.-R. Li \& J.-M. Wang(2025)Li \& Wang}]{Li_2025_Response}
Li, Y.-R., \& Wang, J.-M. 2025, \bibinfo{title}{Radial-dependent Responsivity of Broad-line Regions in Active Galactic Nuclei: Observational Consequences for Reverberation Mapping and Black Hole Mass Measurements,} The Astrophysical Journal, 979, 126, \dodoi{10.3847/1538-4357/ad9fee}

\bibitem[{Y.-R. {Li} {et~al.}(2025){Li}, {Shangguan}, {Wang}, {Davies}, {Santos}, {Eisenhauer}, {Songsheng}, {Winkler}, {Aceituno}, {Bai}, {Bai}, {Brotherton}, {Cao}, {Chen}, {Du}, {Fang}, {Feng}, {Feuchtgruber}, {F{\"o}rster Schreiber}, {Fu}, {Genzel}, {Gillessen}, {Ho}, {Hu}, {Liu}, {Lutz}, {Ott}, {Petrov}, {Rabien}, {Shimizu}, {Sturm}, {Tacconi}, {Wang}, {Yao}, {Zhai}, {Zhang}, {Zhao}, \& {Zhao}}]{SARM_2025}
{Li}, Y.-R., {Shangguan}, J., {Wang}, J.-M., {et~al.} 2025, \bibinfo{title}{{Spectroastrometry and Reverberation Mapping of Active Galactic Nuclei. II. Measuring Geometric Distances and Black Hole Masses of Four Nearby Quasars},} arXiv e-prints, arXiv:2502.18856.
\newblock \doarXiv{2502.18856}

\bibitem[{K. Long {et~al.}(2023)Long, Dexter, Cao, Davies, Eisenhauer, Lutz, Santos, Shangguan, Shimizu, \& Sturm}]{Long_2023}
Long, K., Dexter, J., Cao, Y., {et~al.} 2023, \bibinfo{title}{Confronting a Thin Disk-wind Launching Mechanism of Broad-line Emission in Active Galactic Nuclei with GRAVITY Observations of Quasar 3C 273,} The Astrophysical Journal, 953, 184, \dodoi{10.3847/1538-4357/ace4bb}

\bibitem[{R.~R. {Ludwig} {et~al.}(2012){Ludwig}, {Greene}, {Barth}, \& {Ho}}]{2Component_NLR_Ludwig_Greene}
{Ludwig}, R.~R., {Greene}, J.~E., {Barth}, A.~J., \& {Ho}, L.~C. 2012, \bibinfo{title}{{Physical Properties of the Narrow-line Region of Low-mass Active Galaxies},} \apj, 756, 51, \dodoi{10.1088/0004-637X/756/1/51}

\bibitem[{N. {Murray} \& J. {Chiang}(1997){Murray} \& {Chiang}}]{CM97}
{Murray}, N., \& {Chiang}, J. 1997, \bibinfo{title}{{Disk Winds and Disk Emission Lines},} \apj, 474, 91, \dodoi{10.1086/303443}

\bibitem[{S. Nagoshi {et~al.}(2024)Nagoshi, Iwamuro, Yamada, Ueda, Oikawa, Otsuka, Isogai, \& Mineshige}]{2Component_Nagoshi2024_CSQ}
Nagoshi, S., Iwamuro, F., Yamada, S., {et~al.} 2024, \bibinfo{title}{Probing the origin of the two-component structure of broad-line region by reverberation mapping of an extremely variable quasar,} Monthly Notices of the Royal Astronomical Society, 529, 393, \dodoi{10.1093/mnras/stae319}

\bibitem[{A. {Pancoast} {et~al.}(2011){Pancoast}, {Brewer}, \& {Treu}}]{Pancoast2011}
{Pancoast}, A., {Brewer}, B.~J., \& {Treu}, T. 2011, \bibinfo{title}{{Geometric and Dynamical Models of Reverberation Mapping Data},} \apj, 730, 139, \dodoi{10.1088/0004-637X/730/2/139}

\bibitem[{A. {Pancoast} {et~al.}(2014){Pancoast}, {Brewer}, \& {Treu}}]{PANCOAST_CLOUDS_14}
{Pancoast}, A., {Brewer}, B.~J., \& {Treu}, T. 2014, \bibinfo{title}{{Modelling reverberation mapping data - I. Improved geometric and dynamical models and comparison with cross-correlation results},} \mnras, 445, 3055, \dodoi{10.1093/mnras/stu1809}

\bibitem[{B. Peterson(2006)Peterson}]{Peterson2006}
Peterson, B. 2006, in Physics of Active Galactic Nuclei at all Scales, ed. D.~Alloin, R.~Johnson, \& P.~Lira (Berlin, Heidelberg: Springer Berlin Heidelberg), 77--100, \dodoi{10.1007/3-540-34621-X_3}

\bibitem[{B. {Peterson}(2013){Peterson}}]{STORM_PROP}
{Peterson}, B. 2013, \bibinfo{title}{{Mapping the AGN Broad Line Region by Reverberation},}, HST Proposal ID 13330. Cycle 21

\bibitem[{B.~M. {Peterson}(1993){Peterson}}]{Peterson93_RM}
{Peterson}, B.~M. 1993, \bibinfo{title}{{Reverberation Mapping of Active Galactic Nuclei},} \pasp, 105, 247, \dodoi{10.1086/133140}

\bibitem[{B.~M. {Peterson}(2014){Peterson}}]{Peterson2014_REVIEW}
{Peterson}, B.~M. 2014, \bibinfo{title}{{Measuring the Masses of Supermassive Black Holes},} \ssr, 183, 253, \dodoi{10.1007/s11214-013-9987-4}

\bibitem[{G.~B. {Rybicki} \& D.~G. {Hummer}(1983){Rybicki} \& {Hummer}}]{RH83}
{Rybicki}, G.~B., \& {Hummer}, D.~G. 1983, \bibinfo{title}{{The specific luminosity of a three-dimensional medium in terms of the escape probability},} \apj, 274, 380, \dodoi{10.1086/161454}

\bibitem[{D.~J.~D. {Santos} {et~al.}(2025){Santos}, {Shimizu}, {Davies}, {Cao}, {Dexter}, {de Zeeuw}, {Eisenhauer}, {F{\"o}rster-Schreiber}, {Feuchtgruber}, {Genzel}, {Gillessen}, {Kuhn}, {Lutz}, {Ott}, {Rabien}, {Shangguan}, {Sturm}, \& {Tacconi}}]{2Component_GRAVITY}
{Santos}, D.~J.~D., {Shimizu}, T., {Davies}, R., {et~al.} 2025, \bibinfo{title}{{Spectroscopic AGN survey at $z$ $\sim$ 2 with NTT/SOFI for GRAVITY+ observations},} arXiv e-prints, arXiv:2503.02942, \dodoi{10.48550/arXiv.2503.02942}

\bibitem[{V.~V. {Sobolev}(1957){Sobolev}}]{Sobolev}
{Sobolev}, V.~V. 1957, \bibinfo{title}{{The Diffusion of L{\ensuremath{\alpha}} Radiation in Nebulae and Stellar Envelopes.},} \sovast, 1, 678

\bibitem[{T. {Waters} {et~al.}(2016){Waters}, {Kashi}, {Proga}, {Eracleous}, {Barth}, \& {Greene}}]{WATERS16}
{Waters}, T., {Kashi}, A., {Proga}, D., {et~al.} 2016, \bibinfo{title}{{Reverberation Mapping of the Broad Line Region: Application to a Hydrodynamical Line-driven Disk Wind Solution},} \apj, 827, 53, \dodoi{10.3847/0004-637X/827/1/53}

\bibitem[{W.~F. {Welsh} \& K. {Horne}(1991){Welsh} \& {Horne}}]{Welsh&Horne_EchoMaps}
{Welsh}, W.~F., \& {Horne}, K. 1991, \bibinfo{title}{{Echo Images of Broad-Line Regions in Active Galactic Nuclei},} \apj, 379, 586, \dodoi{10.1086/170530}

\bibitem[{P.~R. {Williams} {et~al.}(2020){Williams}, {Pancoast}, {Treu}, {Brewer}, {Peterson}, {Barth}, {Malkan}, {De Rosa}, {Horne}, {Kriss}, {Arav}, {Bentz}, {Cackett}, {Dalla Bont{\`a}}, {Dehghanian}, {Done}, {Ferland}, {Grier}, {Kaastra}, {Kara}, {Kochanek}, {Mathur}, {Mehdipour}, {Pogge}, {Proga}, {Vestergaard}, {Waters}, {Adams}, {Anderson}, {Ar{\'e}valo}, {Beatty}, {Bennert}, {Bigley}, {Bisogni}, {Borman}, {Boroson}, {Bottorff}, {Brandt}, {Breeveld}, {Brotherton}, {Brown}, {Brown}, {Canalizo}, {Carini}, {Clubb}, {Comerford}, {Corsini}, {Crenshaw}, {Croft}, {Croxall}, {Deason}, {De Lorenzo-C{\'a}ceres}, {Denney}, {Dietrich}, {Edelson}, {Efimova}, {Ely}, {Evans}, {Fausnaugh}, {Filippenko}, {Flatland}, {Fox}, {Gardner}, {Gates}, {Gehrels}, {Geier}, {Gelbord}, {Gonzalez}, {Gorjian}, {Greene}, {Grupe}, {Gupta}, {Hall}, {Henderson}, {Hicks}, {Holmbeck}, {Holoien}, {Hutchison}, {Im}, {Jensen}, {Johnson}, {Joner}, {Jones}, {Kaspi}, {Kelly}, {Kennea}, {Kim}, {Kim}, {Kim}, {King}, {Klimanov}, {Knigge},
  {Krongold}, {Lau}, {Lee}, {Leonard}, {Li}, {Lira}, {Lochhaas}, {Ma}, {MacInnis}, {Manne-Nicholas}, {Mauerhan}, {McGurk}, {McHardy}, {Montuori}, {Morelli}, {Mosquera}, {Mudd}, {M{\"u}ller-S{\'a}nchez}, {Nazarov}, {Norris}, {Nousek}, {Nguyen}, {Ochner}, {Okhmat}, {Papadakis}, {Parks}, {Pei}, {Penny}, {Pizzella}, {Poleski}, {Pott}, {Rafter}, {Rix}, {Runnoe}, {Saylor}, {Schimoia}, {Scott}, {Sergeev}, {Shappee}, {Shivvers}, {Siegel}, {Simonian}, {Siviero}, {Skielboe}, {Somers}, {Spencer}, {Starkey}, {Stevens}, {Sung}, {Tayar}, {Tejos}, {Turner}, {Uttley}, {Van Saders}, {Vaughan}, {Vican}, {Villanueva}, {Villforth}, {Weiss}, {Woo}, {Yan}, {Young}, {Yuk}, {Zheng}, {Zhu}, \& {Zu}}]{STORM_CLOUDS}
{Williams}, P.~R., {Pancoast}, A., {Treu}, T., {et~al.} 2020, \bibinfo{title}{{Space Telescope and Optical Reverberation Mapping Project. XII. Broad-line Region Modeling of NGC 5548},} \apj, 902, 74, \dodoi{10.3847/1538-4357/abbad7}

\bibitem[{X.-G. Zhang(2011)Zhang}]{2Component_Zhang_2011_ILR}
Zhang, X.-G. 2011, \bibinfo{title}{EVIDENCE FOR THE INTERMEDIATE BROAD-LINE REGION OF REVERBERATION-MAPPED ACTIVE GALACTIC NUCLEUS PG 0052+251,} The Astrophysical Journal, 741, 104, \dodoi{10.1088/0004-637X/741/2/104}

\bibitem[{Z.-X. {Zhang} {et~al.}(2019){Zhang}, {Du}, {Smith}, {Zhao}, {Hu}, {Xiao}, {Li}, {Huang}, {Wang}, {Bai}, {Ho}, \& {Wang}}]{RM3c27319}
{Zhang}, Z.-X., {Du}, P., {Smith}, P.~S., {et~al.} 2019, \bibinfo{title}{{Kinematics of the Broad-line Region of 3C 273 from a 10 yr Reverberation Mapping Campaign},} \apj, 876, 49, \dodoi{10.3847/1538-4357/ab1099}

\bibitem[{L. Zhu {et~al.}(2009)Zhu, Zhang, \& Tang}]{2Component_Zhu_2009_ILR_VBLR}
Zhu, L., Zhang, S.~N., \& Tang, S. 2009, \bibinfo{title}{EVIDENCE FOR AN INTERMEDIATE LINE REGION IN ACTIVE GALACTIC NUCLEI's INNER TORUS REGION AND ITS EVOLUTION FROM NARROW TO BROAD LINE SEYFERT I GALAXIES,} The Astrophysical Journal, 700, 1173, \dodoi{10.1088/0004-637X/700/2/1173}

\end{thebibliography}
\bibliographystyle{aasjournal}

\appendix 
\vspace{-1mm}
\section{Fitting results} \label{appendix}
This paper began with the idea that, given that we had successfully fit BLR interferometry data with a disk-wind type model, we should similarly be able to fit such a model to reverberation mapping data. We picked the AGN-STORM dataset as at the time of inception it was highest quality reverberation mapping dataset (now there is a second AGN-STORM experiment available, and a ground-based velocity-resolved campaign of 3C 273 offers similar quality as well). We started by fitting our simple disk-wind model to the line profile produced a synthetic lightcurve of the integrated emission line (a 1D RM fit) to compare to the observations. Doing this works of similar quality to previous work, as Figure \ref{fig:LCfit} below shows:
\begin{figure*}[!ht]
    \begin{minipage}{\textwidth}
    \centering
    \includegraphics[width=1.\linewidth,keepaspectratio]{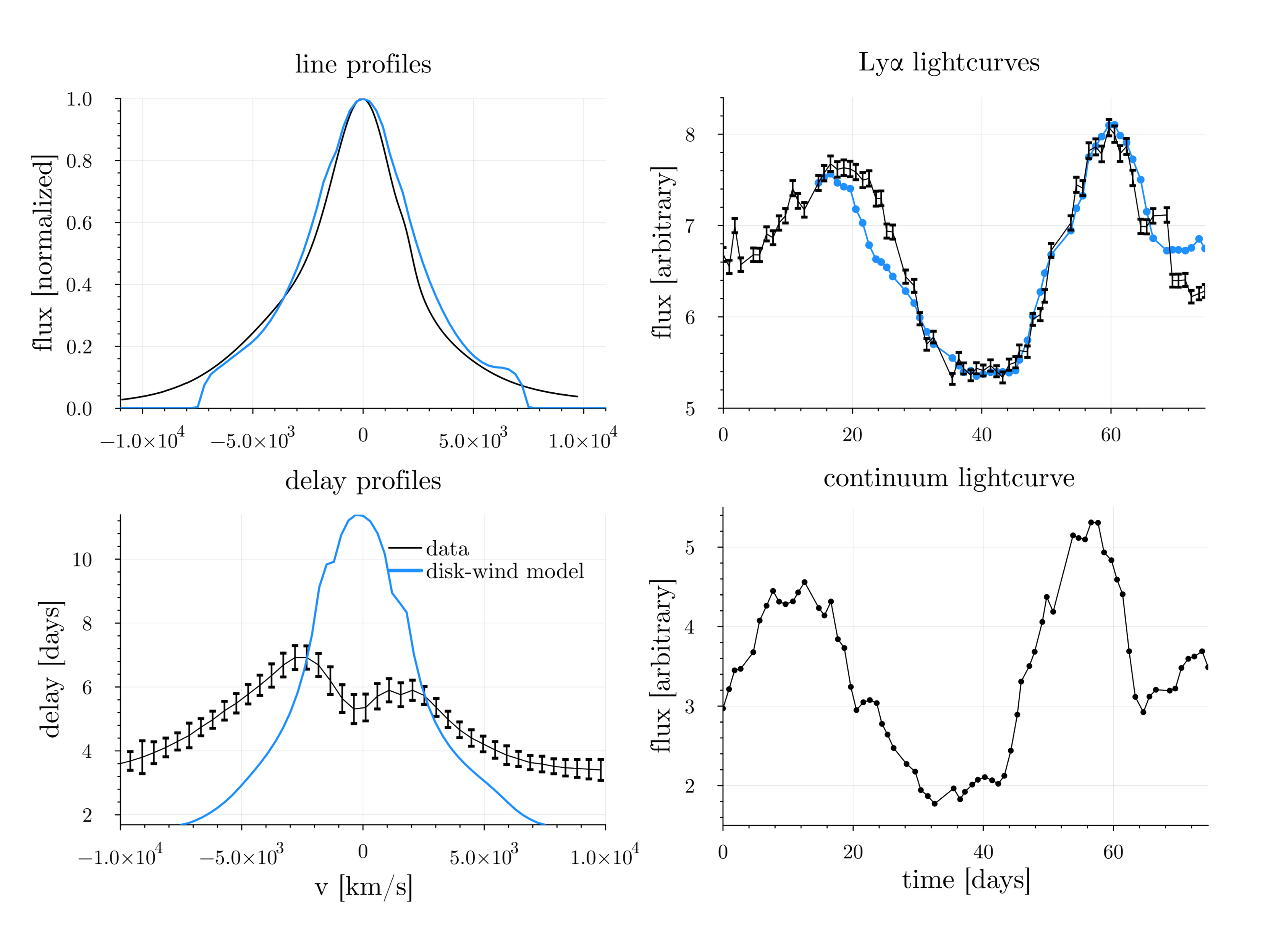}
    \caption{A sample line profile and synthetic lightcurve for a disk-wind model fit to the AGN-STORM data. While the line profile fit is poor in the wings, note the overall structure and in particular the resulting lightcurve are qualitatively very similar in fit quality to previous cloud fitting results, i.e. the integrated lightcurve shown in \cite{STORM_CLOUDS}. Clearly the delay profile is wrong and has just a single peak for reasons discussed in the main text, but note it is also qualitatively similar to the previous cloud-fitting delay profiles presented in \cite{STORM_CLOUDS}.}   
    \label{fig:LCfit}
    \end{minipage}
\end{figure*}

However the STORM data is so remarkable and precise that we also wanted to match the velocity resolved delay profile across the line, but were surprised to find that our model could not match this. We ran a separate experiment and told the model to explicitly fit for this quantity, and while it was able to roughly do so the line profile all such fits converged to was much too broad for the data. This motivated the problem discussed in section \ref{sec:problem}. An example fit from this procedure is shown below in Figure \ref{fig:RMfit}:

\begin{figure*}[!ht]
    \begin{minipage}{\textwidth}
    \centering
    \includegraphics[width=1.\linewidth,keepaspectratio]{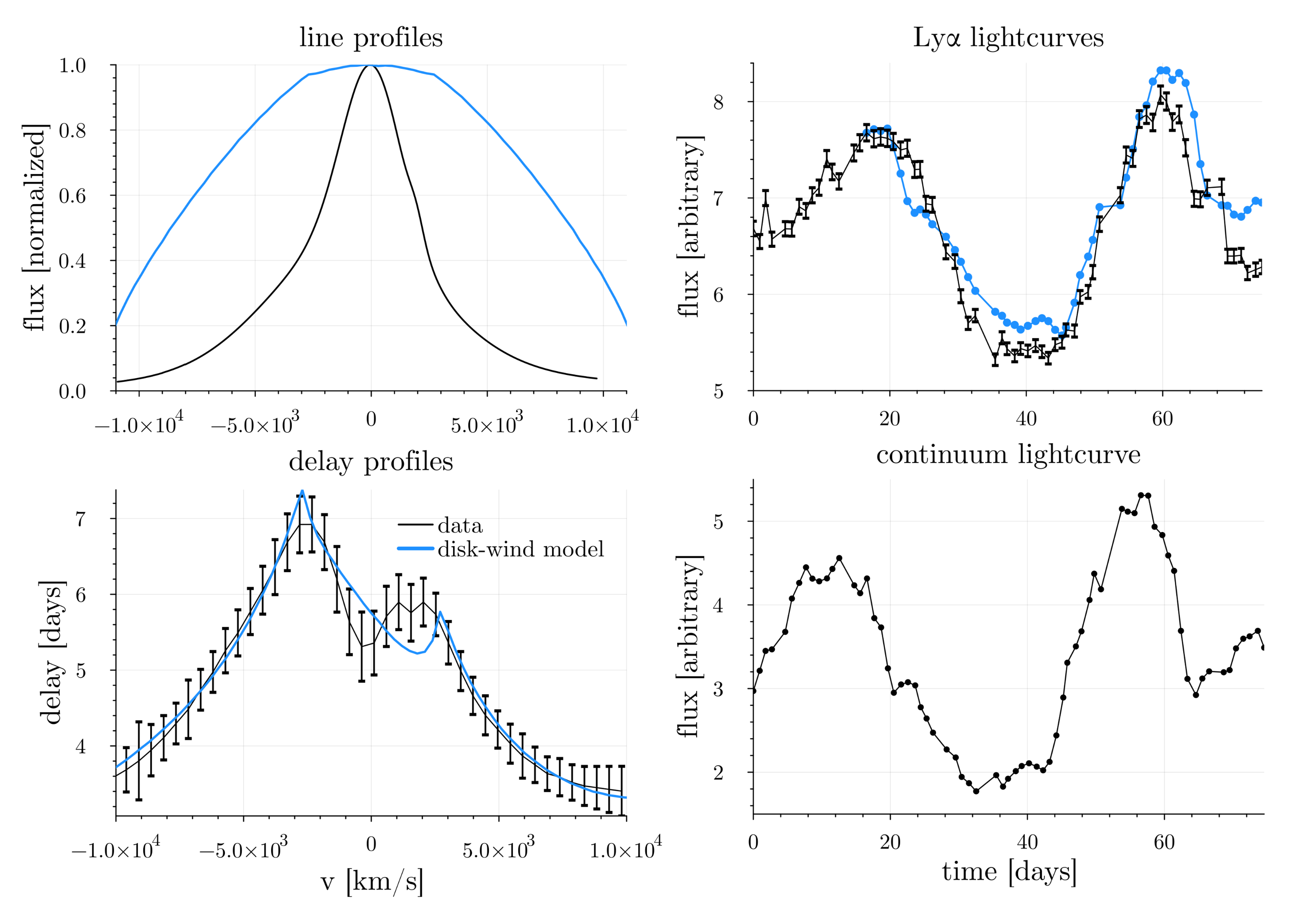}
    \caption{A sample line profile and synthetic lightcurve for a disk-wind model fit to the AGN-STORM velocity-resolved delay profile data. Note that while the lightcurve fit is still qualitatively good, and the delay profile shape now matches the data much more closely, the line profile is much too broad and clearly a very poor fit for reasons discussed in the main text.}
    \label{fig:RMfit}
    \end{minipage}
\end{figure*}

The literature shows that often cloud models produce single peaks in the delays \citep[i.e.][]{STORM_CLOUDS,PANCOAST_CLOUDS_14}, further confounding the problem. We attempted to fit a version of the cloud model to the data and confirmed this problem and found similar results\textemdash the clouds can produce two peaks in the delay profile if they are sufficiently ordered and disk-like, but doing so ruins the line-profile shape as discussed in the main text. 

Thus if we look at the 1D quantities only, we can ``fit" the data roughly as well with a disk-wind model as others have with clouds, but doing this ignores the full 2D STORM data products and thus does not match the double-peaked morphology seen in C~{\sc iv} and Ly$\alpha$ delay profiles. We can fit the line profile or the delay profile (more roughly, see main text) with either single-component model, but not both simultaneously, as the combination of choices that produce the observed delay profiles in the model result in line profiles that do not match and vice versa.

Interestingly, when fitting for one product vs the other they also prefer different physical parameters, namely in the inclination angle, black hole mass, and power law dependence of the source function (in the disk-wind case). To match the delay profile requires a steeper drop off in flux in the disk (running up against our prior boundary of $\alpha = 2$, compared with matching the line profile alone which prefers $\alpha \approx 0.5$), a larger black hole mass (roughly $2\times$ the value when fitting for the line profile), and a higher inclination angle ($\sim 45^\circ$ vs $\sim 20^\circ$ for the delay profile and line profile fits, respectively). 

This motivated our exploration of the three cases described in the main text. The first case involves simply blocking off the far side of the system to shorten the delays, and is thus not much more complicated than either the standard cloud or disk-wind model of the BLR. Thus we attempted to fit this class of model to the data, but still could not properly match the data products. The best-fits from this experiment are shown below in Figure \ref{fig:bestDelayFits}, illustrating the problem described in the main text with the width/amplitude of the delay/line profiles being difficult to explain with a single-component model:
\begin{wrapfigure}{L}{0.45\textwidth}
    \centering
    \includegraphics[width=1.\linewidth,keepaspectratio]{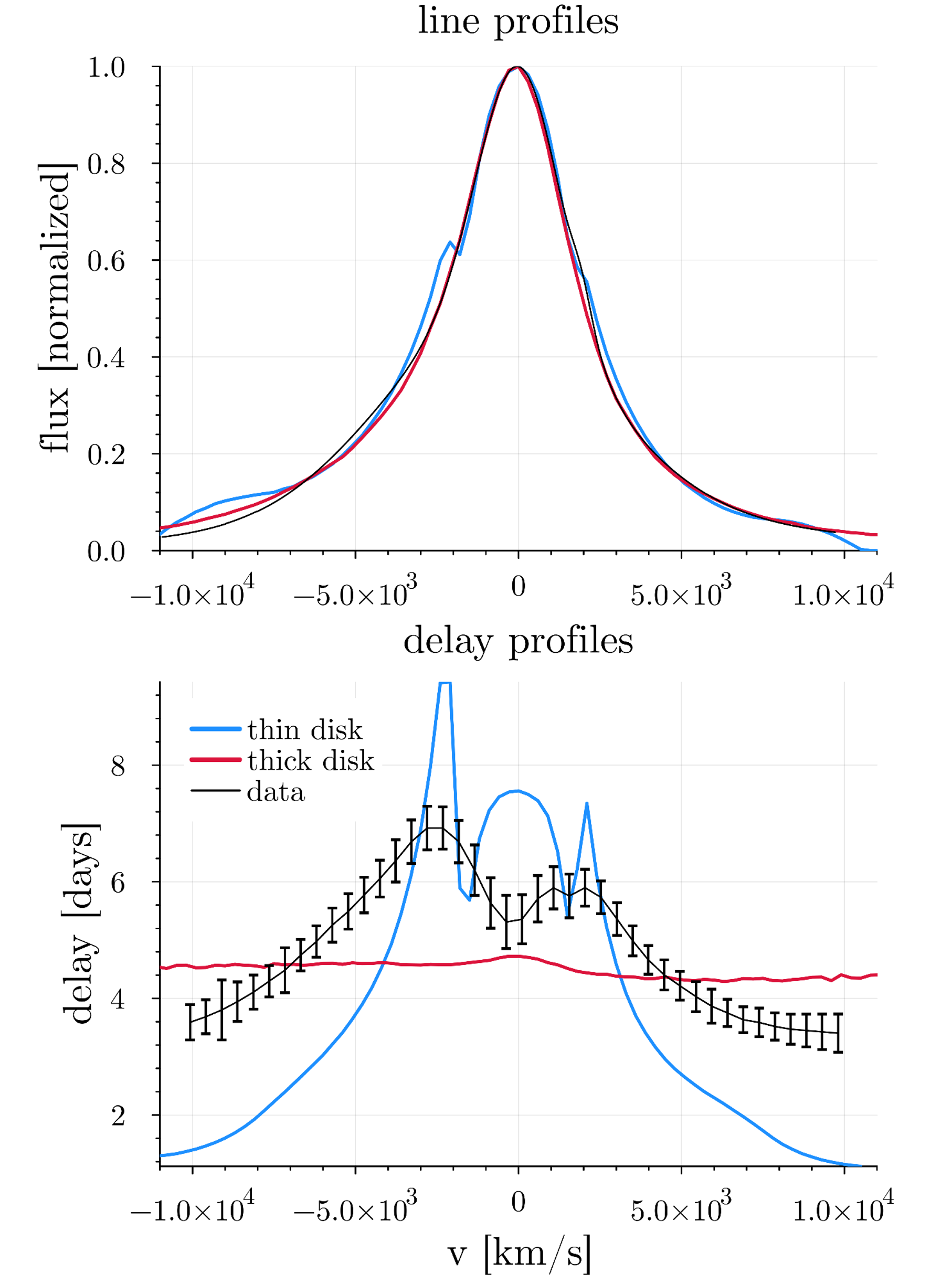}
    \caption{Best fits of ``\hyperref[case1]{case 1}" style models (single component cloud/thick disk and disk-wind models where a portion of the system can be obscured). Notice that while both can match the line profile relatively well, the delay profile is poorly fit in either case due to the problems discussed in the main text.}
    \label{fig:bestDelayFits}
\end{wrapfigure}

There are many possible parameter combinations in either a disk-wind or cloud type model that can be varied which we have not fully explored\textemdash just because our attempts at sampling the distributions both by hand and numerically have not found a satisfactory combination of parameters that work does not necessarily mean one does not exist. Nevertheless we think the computational methods we have tried are relatively robust, and if such a place exists in the parameter space it must be small, specific, and thus may be relatively contrived as a result, leading us to disfavor this option as a method for explaining the data.

\end{document}